\documentclass[aps,prl,reprint,noeprint,superscriptaddress]{revtex4-2}

\usepackage{graphicx}
\usepackage[colorlinks]{hyperref}
\usepackage{natbib}
\usepackage{bm}
\usepackage{mathtools}
\usepackage{color}
\usepackage{bbold}
\usepackage{lipsum}
\usepackage[utf8]{inputenc}
\newcommand{\rev}[1]{\textcolor{black}{#1}}
\renewcommand{\Im}{\mathrm{Im}}
\renewcommand{\Re}{\mathrm{Re}}
\newcommand{\norm}[1]{\left \lVert #1 \right \rVert}

\usepackage{bm}

\renewcommand{\matrix}{}
\renewcommand{\vec}{\boldsymbol}

\begin{document}
%TC:ignore

%Title of paper
\title{Detecting and Focusing on a Nonlinear Target in a Complex Medium}

% repeat the \author .. \affiliation  etc. as needed
% \email, \thanks, \homepage, \altaffiliation all apply to the current
% author. Explanatory text should go in the []'s, actual e-mail
% address or url should go in the {}'s for \email and \homepage.
% Please use the appropriate macro foreach each type of information

% \affiliation command applies to all authors since the last
% \affiliation command. The \affiliation command should follow the
% other information
% \affiliation can be followed by \email, \homepage, \thanks as well.
\author{Antton Goïcoechea}
\email[]{antton.goicoechea@univ-rennes.fr}
\affiliation{Université de Rennes, CNRS, IETR - UMR 6164; F-35000 Rennes, France}

\author{Jakob Hüpfl}
% \email[]{}
\affiliation{Institute for Theoretical Physics, Vienna University of Technology (TU Wien),  A-1040 Vienna, Austria}

\author{Stefan Rotter}
% \email[]{stefan.rotter@tuwien.ac.at}
\affiliation{Institute for Theoretical Physics, Vienna University of Technology (TU Wien),  A-1040 Vienna, Austria}

\author{François Sarrazin}
% \email[]{francois.sarrazin@univ-rennes.fr}
\affiliation{Université de Rennes, CNRS, IETR - UMR 6164; F-35000 Rennes, France}

\author{Matthieu Davy}
\email[]{matthieu.davy@univ-rennes.fr}
\affiliation{Université de Rennes, CNRS, IETR - UMR 6164; F-35000 Rennes, France}

\date{\today}

\begin{abstract}
Wavefront shaping techniques allow waves to be focused on a diffraction-limited target deep inside disordered media. To identify the target position, a guidestar is required that typically emits a frequency-shifted signal. Here we present a noninvasive matrix approach operating at a single frequency only, based on the variation of the field scattered by a nonlinear target illuminated at two different incident powers. The local perturbation induced by the nonlinearity serves as a guide for identifying optimal incident wavefronts. We demonstrate maximal focusing on electronic devices embedded in chaotic microwave cavities and extend our approach to temporal signals. Finally, we exploit the programmability offered by reconfigurable smart surfaces to enhance the intensity delivered to a nonlinear target. Our results pave the way for deep imaging protocols that use any type of nonlinearity as feedback, requiring only the measurement of a monochromatic scattering matrix.
\end{abstract}

% insert suggested keywords - APS authors don't need to do this
%\keywords{}

%\maketitle must follow title, authors, abstract, and keywords
\maketitle
%TC:endignore

\noindent \textit{Introduction.---}Wavefront shaping techniques can partially counteract the effect of disorder by coherently controlling wave-matter interaction \cite{moskControllingWavesSpace2012,rotterLightFieldsComplex2017,caoShapingPropagationLight2022}. Of particular interest is the possibility to focus waves on a diffraction-limited focal spot inside or behind a strongly scattering medium~\cite{vellekoopExploitingDisorderPerfect2010} or to deposit energy to a target region ~\cite{jeongFocusingLightEnergy2018,benderDepthtargetedEnergyDelivery2022}. In the linear regime, when the field at the target location is directly accessible, the incident wavefront for focusing can be optimally tailored in space and/or time using techniques such as phase-conjugation for monochromatic waves~\cite{yaqoobOpticalPhaseConjugation2008,popoffMeasuringTransmissionMatrix2010,xuTimereversedUltrasonicallyEncoded2011,hsuCorrelationenhancedControlWave2016}, time reversal for broadband signals~\cite{derodeRobustAcousticTime1995} or the eigenstates of an operator constructed from the scattering matrix~\cite{kimMaximalEnergyTransport2012,lambertDistortionMatrixApproach2020,benderDepthtargetedEnergyDelivery2022}. However, embedding a detector within a scattering medium is an invasive procedure that rules out many applications in deep optical imaging, wireless communications, wireless power transfer, or sensing. Noninvasive approaches therefore rely on the presence of a guidestar within the medium~\cite{xuTimereversedUltrasonicallyEncoded2011,judkewitzSpecklescaleFocusingDiffusive2013,laiPhotoacousticallyGuidedWavefront2015,horstmeyerGuidestarassistedWavefrontshapingMethods2015,caoShapingPropagationLight2022}. 

The nonlinearity of wave-matter interaction has emerged in this context as an efficient approach for deep imaging. Nonlinear techniques rely on the localized feedback generated by a nonlinear target. In acoustics, microbubbles serve as contrast agents for ultrasound  imaging~\cite{millerUltrasonicDetectionResonant1981,qinUltrasoundContrastMicrobubbles2009} while in optics, Raman microscopy~\cite{evansCoherentAntiStokesRaman2008}, two-photon fluorescence~\cite{katzNoninvasiveNonlinearFocusing2014} or second-harmonic generation~\cite{deaguiarEnhancedNonlinearImaging2016, moonMeasuringScatteringTensor2023} have been exploited to obtain a diffraction-limited focal spot~\cite{horstmeyerGuidestarassistedWavefrontshapingMethods2015,yoonDeepOpticalImaging2020}.
In the microwave regime, most electronic devices, even as simple as a diode~\cite{reisnerSelfshieldedTopologicalReceiver2020,jeonNonHermitianSymmetricSpectral2020,suwunnaratNonlinearCoherentPerfect2022}, exhibit a nonlinear behavior and can be detected by harmonic radars in cluttered environments~\cite{mazzaroNonlinearRadarFinding2017,perezDetectingPresenceElectronic2022}.
All these techniques nevertheless require complex experimental setups to detect and/or filter the frequency-shifted nonlinear signal.

A local perturbation of a linear scattering medium can also serve as a guidestar~\cite{ambichlFocusingDisorderedMedia2017,horodynskiOptimalWaveFields2020,delhougneCoherentWaveControl2021,bouchetMaximumInformationStates2021,bouchetTemporalShapingWave2023,yeoTimeReversalCommunications2022,solOptimalBlindFocusing2024}.
Any change within a disordered sample is encoded in the random speckle pattern resulting from the complex interaction of the incident wave and the sample~\cite{berkovitsSensitivityMultiplescatteringSpeckle1991}.
Therefore, the derivative of the scattering matrix $S(\omega)$ with respect to a parameter $\theta$, i.e. $\partial_\theta S$, contains information about $\theta$.
For unitary scattering matrices, the eigenvalues of the generalized Wigner-Smith (WS) operator $Q = -iS^{-1} \partial_\theta S$ indicate how strongly the conjugate quantity to $\theta$ is affected by a variation~\cite{ambichlFocusingDisorderedMedia2017,horodynskiOptimalWaveFields2020,orazbayev2024wave}.
The operator $(\partial_\theta S)^\dagger \partial_\theta S$ also turns out to measure the content of Fisher information carried by the scattered wave on the parameter $\theta$~\cite{bouchetMaximumInformationStates2021,horodynskiInvariancePropertyFisher2021,hupflContinuityEquationFlow2024}.
The eigenstates of these operators thus provide the solution for maximal focusing, micromanipulation or for optimal sensitivity with respect to  $\theta$. However, setting up these operators requires a variation of the target parameter(s) in the first place, which is hard to accomplish in linear static systems without invasive external intervention inside the scattering medium.

Here, we present a noninvasive approach for optimal focusing on a nonlinear target in static scattering systems. Most importantly, our approach does not require a measurement of {a higher-}harmonic response of the target device  {or any other frequency-shifted signal.} This allows for the detection of all types of nonlinearities {(whether they induce $n$-th order harmonic generation, or a Kerr effect for example)}, and it does not rely on prior knowledge on the medium or on the target. All these requirements are satisfied by leveraging the nonlinear scattering response, which we probe by tuning the incident power~\cite{flemingPerturbationTransmissionMatrices2019,moonMeasuringScatteringTensor2023}.
%Specifically, we extract wavefronts focusing in space and time by measuring the response of the system for two incident powers and by filtering out the expected linear response. Theoretical results are confirmed by microwave measurements within scattering systems. Finally, we show that the intensity focused on the nonlinear target can be further enhanced in programmable environments by optimizing reconfigurable metasurfaces.

\begin{figure*}
\centering
\includegraphics[width=2.05\columnwidth]{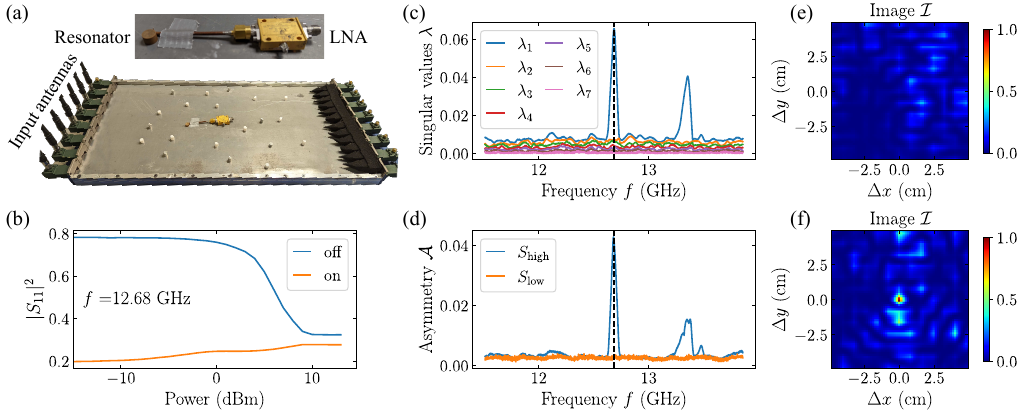}
\caption{\label{fig: exmpt 2D 1 target} (a) Photograph of the experimental setup. A resonator coupled in the near field to a wire antenna connected to a (nonlinear) low-noise amplifier (LNA) is located within a two-dimensional multimode waveguide (cover plate not shown). The scattering matrix is measured using seven antennas at the left interface. An absorbing foam is placed at the right interface to mimic open boundary conditions. (b) Reflection parameter $r = |S_{11}|^2$ of the LNA as a function of incident power $P$ for an empty single-mode waveguide at $f_0 = 12.68$~GHz with the second port connected to the LNA (powered --`on' -- or not -- `off') terminated by an open-circuit condition.
(c,d) Spectra of the singular values $\lambda_n$ of $\Delta S$ (c) and asymmetry factor $\mathcal{A}$ (d). The nonlinearity is maximal at the resonance of the nonlinear target $f = f_0$ indicated by a black-dashed line. (e,f) Intensity map within the system for a random illumination (e) or with $\vec{c}^\text{in} = \vec{d}^*$ (f). Both maps are normalized by the maximum value obtained for the optimized wavefront.
}
\end{figure*}

\noindent \textit{Theoretical analysis.---}We consider a system made up of dielectric obstacles described by the dielectric function $\epsilon(\vec{r})$, which we probe by coherent monochromatic electromagnetic waves at frequency $\omega$.
%{ Our objective is to find an incident field $\vec{E}_{\omega}^\text{in}$ that creates an optimal focus on a nonlinear target described by the polarization $\vec{P}^\text{NL}(\vec{E})$, which is nonlinear in the electric field $\vec{E}$ embedded inside this scattering environment (e.g. $\vec{D}(\vec{E}) = \epsilon_0 \epsilon \vec{E} + \vec{P}^\text{NL}(\vec{E})$). }
{Our objective is to find an incident field $\vec{E}_{\omega}^\text{in}$ that creates an optimal focus on a target described by the polarization $\vec{P}^\text{NL}(\vec{E})$, which is nonlinear in the electric field $\vec{E}$, embedded inside this scattering environment.}

{To disentangle the linear (L) from the nonlinear (NL) scattering response, we write the far-field solution $\vec{E}_{\omega}^\alpha(\vec{r})$ at frequency $\omega$ for an incident field strength $\alpha$ using the following exact integral equation: 
\begin{equation}
\label{eq:integral_eq}
    \vec{E}_{\omega}^\alpha(\vec{r}) = \alpha \vec{E}_{\omega}^\text{L}(\vec{r}) + \frac{k^2}{\epsilon_0} \int_{\mathbb{R}^3} d\vec{r}\,' G_\omega(\vec{r},\vec{r}\,') \vec{P}_{\omega}^{\mathrm{NL}}[\vec{E}^\alpha(\vec{r}\,')].
\end{equation}
Here, $\vec{E}_{\omega}^\text{L}(\vec{r})$ describes the linear component of the field for normalized amplitude $\alpha = 1$ and $G_{\omega}(\vec{r},\vec{r}\,')$ is the Green's tensor of the system without nonlinearities. The target's nonlinear polarization field component at frequency $\omega$ is given by $\vec{P}_{\omega}^{\mathrm{NL}}[\vec{E}^\alpha(\vec{r}')]$ and depends in general on all frequency components of the electric field $\vec{E}^\alpha_{\omega_1}(\vec{r}'),\vec{E}^\alpha_{\omega_2}(\vec{r}'), \dots$ at $\vec{r}'$. However, for easier notation we write this in Eq.~\eqref{eq:integral_eq} as a dependence on the full field $\vec{E}^\alpha(\vec{r}) = \int_0^\infty \Re(\vec{E}_{\omega}^\alpha(\vec{r}) e^{-i\omega t}) d \omega $.}

%Note that both the linear response (the first term) of the system and the nonlinear response (the second term) are considered to be time-harmonic waves oscillating at the same frequency. 
The idea is now to eliminate the linear term and to time-reverse the nonlinear signal to create a focus at the target (see Supplementary Material~\cite{supp_mat} for details). We achieve this by varying the amplitude of the incident field by $\Delta \alpha$, then the difference of the amplitude-normalized fields $\delta \vec{E}_{\omega}(\vec{r}) = (\alpha+\Delta \alpha)^{-1} \vec{E}_{\omega}^{\alpha+\Delta \alpha}(\vec{r}) - \alpha^{-1} \vec{E}_{\omega}^\alpha(\vec{r})$ satisfies 
\begin{equation}
    \delta \vec{E}_{\omega}(\vec{r}) = \frac{k^2}{\epsilon_0} \int_{\mathbb{R}^3} d\vec{r}\,' G_{\omega}(\vec{r},\vec{r}\,') \delta \vec{P}_{\omega}^{\mathrm{NL}}(\vec{r}\,')\,.
\label{Eq:dE}
\end{equation}
\noindent We see that {$\delta \vec{P}_{\omega}^{\mathrm{NL}}(\vec{r}\,') = (\alpha + \Delta \alpha)^{-1}\vec{P}_{\omega}^{\mathrm{NL}}[\vec{E}^{\alpha + \Delta \alpha}(\vec{r}\,')] - \alpha^{-1} \vec{P}_{\omega}^{\mathrm{NL}}[\vec{E}^\alpha(\vec{r}\,')]$}
acts as a source term for $\delta \vec{E}_{\omega}(r)$ with the coupling from the polarization field to the far field being given by the Green's tensor $G_{\omega}(\vec{r},\vec{r}\,')$. {Thus, as long as the incident field interacts with the nonlinearity, the time-reversed of $\delta \vec{E}(\vec{r})$ provides a focus from the far field onto the nonlinear target.}

This becomes especially apparent for a point-like nonlinearity located at $\vec{r}_0$, where the field difference provides the exact coupling between the location of the nonlinearity to the far field, i.e. $\delta \vec{E}_{\omega}(r) \propto G_{\omega}(\vec{r},\vec{r}_0)$. {If we further assume that only a single electromagnetic mode couples to the nonlinearity, then the focus that is created at $\vec{r}_0$ with this method is necessarily optimal for a given incident strength $\alpha$.}
%This is the case for nonlinearities that are connected to the system through single polarization antennas or in systems with only a single polarization degree of freedom, e.g. a thin waveguide as shown in Fig.~\ref{fig: exmpt 2D 1 target} (a).}
Note that throughout the article, `optimal' and `maximal' refer to states obtainable in the linear regime because the nonlinearity is assumed to be weak (perturbative regime).

{From now on, we focus on scalar waves. For linear systems, the linear scattering matrix $\matrix{S}^\text{L}$ connects all input and output channels $\vec{c}^\text{in}$ and $\vec{c}^\text{out}$ through $\vec{c}^\text{out} = \matrix{S}^\text{L} \vec{c}^\text{in}$. For nonlinear systems, the superposition principle and therefore this linear relation no longer hold since the measured output field in $\vec{c}^\text{out}$ must encapsulate the coupling between the incident field emitted in $\vec{c}^\text{in}$ and the nonlinearity. This nonlinear response emanates from the position of the nonlinearity $\vec{r}_0$ and scatters into the output channels quantified by the coefficient vector $\vec{d}$ in accordance with the Green's function $G_{\omega}(\vec{r},\vec{r}_0)$. Our goal is to split the incident field into a linear part that does not interact with the nonlinearity and a nonlinear part. The Green's identity tells us that all incident fields $\vec{c}^\text{in}$ that are orthogonal to the phase conjugate of $\vec{d}$ (i.e. $\vec{d}^*$) have vanishing electric field at the location of the nonlinearity $\vec{E}_{\omega}^\text{L}(\vec{r}_0) = 0$.
This means that the nonlinear contribution to the output field must be a function of $\vec{d} \cdot \vec{c}^\text{in}$:
%This makes it possible to introduce a correction to the relation between the incident and outgoing fields
\begin{equation}
\label{eq:in_out}
    \vec{c}^\text{out} = \matrix{S}^\text{L} \vec{c}^\text{in} + \alpha^{-1} \vec{d} f( \vec{d} \cdot \vec{c}^\text{in}),
\end{equation}
where the nonlinearity of the system is captured by the {arbitrary} nonlinear scalar function $f$ (see Supplementary Material~\cite{supp_mat}).
}

If we now probe the system 
%with basis states $\vec{e}_n$ of incident modes 
at a given input power $\alpha$, the resulting input-output relation can be expressed through the following power-dependent scattering matrix
\begin{equation}
    \matrix{S}^\alpha_{m,n} = \matrix{S}^\text{L}_{m,n} + \alpha^{-1} d_m f(\alpha d_n),
\end{equation}
where $\matrix{S}^\alpha_{m,n}$ corresponds to the output in {channel $m$}
%mode $\vec{e}_m$ 
for an input in {channel $n$}.
%mode $\vec{e}_n$.
However, it is important to note that $\matrix{S}^\alpha$ is not a conventional scattering matrix. While it describes the scattering for the set of incident {fields} at a given power, it cannot encompass the full nonlinear nature of Eq.~\eqref{eq:in_out}. 

Nevertheless, $\matrix{S}^\alpha$ can now be used to identify the optimal focusing {input} by using the difference matrix
\begin{equation}
\begin{split}
    \Delta \matrix{S}_{m,n} =& \matrix{S}_{m,n}^{\alpha + \Delta \alpha} - \matrix{S}_{m,n}^{\alpha} \\
    =&  d_m \left\{ (\alpha+\Delta \alpha)^{-1} f[(\alpha+\Delta \alpha) d_n] -  \alpha^{-1} f(\alpha d_n)\right\}.
\end{split}
\end{equation}
The rank of the matrix is equal to the number of point-like nonlinearities (here one), each coupling to the far field through a unique incident wavefront.
$\vec{d}$ can now be estimated by applying a singular value decomposition (SVD) on $\Delta S = U \lambda V^{*}$. The left singular vector $\vec{U}_1$ of the largest singular value $\lambda_1$ corresponds to $\vec{d}$. By applying a phase-conjugation, the incident wavefront $\vec{d}^*$ provides maximal focusing onto the target. 

The nonlinear scattering coefficient of the target depends on the field $E_n(r_0)$ transmitted by channel $n$. In the absence of global symmetries, we have $E_n(r_0) \neq E_m(r_0)$ in general for $n \neq m$. In contrast to linear systems with modulation of an antenna impedance or dielectric permittivity of a subwavelength object~\cite{horodynskiOptimalWaveFields2020,delhougneCoherentWaveControl2021,yeoTimeReversalCommunications2022,solOptimalBlindFocusing2024}, the right and left singular vector of $\Delta S$ are therefore not equal.

\noindent \textit{Experimental results.---}In our first experiment, we consider a nonlinear target embedded within a two-dimensional scattering system working in the microwave range. {The two-dimensional geometry is chosen so that the field can be measured `inside' the system.} The target consists of a high-Q dielectric cylinder with a refraction index of $n \approx 6$ coupled in its near field to a wire antenna connected to a low-noise amplifier (LNA). This nonlinear passive target - the LNA is not powered - is conceptually similar to the coupling of a resonator to a short-circuited diode found in Refs~\cite{reisnerSelfshieldedTopologicalReceiver2020,jeonNonHermitianSymmetricSpectral2020,suwunnaratNonlinearCoherentPerfect2022} for nonlinear coherent perfect absorption or the formation of defect modes. However, we find that our target exhibits a more pronounced nonlinearity due to the presence of transistors in the LNA. This is crucial since the target is excited from its far field. 
We characterize the reflectivity of the LNA by connecting it to a single-mode waveguide whose reflection parameter $|r(\omega)|^2  = |S_{11}|^2$ is measured for increasing input power $P$. While $|r(\omega)|^2$ is constant at low powers indicating a linear behavior, we observe in Fig.~\ref{fig: exmpt 2D 1 target}(b) that $|r(\omega)|^2$ rapidly decreases for $P>0$ dBm before saturating for $P>10$ dBm. The nonlinearity therefore results from enhanced absorption within the LNA at high power.

The nonlinear target is then placed within a multimode waveguide (cavity) supporting a single mode in its vertical direction between 12 GHz and 15 GHz~\cite{davyMeanPathLength2021,delhougneExperimentalRealizationOptimal2021}. Seven metallic and 13 dieletric scatterers are randomly placed inside the cavity to randomize the field. {Placing the nonlinear device in a disordered cavity 
%guarantees that the target receives enough energy to be in the nonlinear regime, and also 
highlights the performance of the approach in cases where the Green's functions within the system are unknown.} The flux-normalized matrix $S^\alpha(\omega)$ is measured using a vector network analyzer (VNA) between $N=7$ single-mode waveguides that are connected to the left interface of the cavity by coax-to-waveguide transitions. At the right interface, we place an absorbing foam to mimic open boundary conditions. The difference matrix $\Delta S(\omega)$ is constructed from measurements of $S^\alpha(\omega)$ at two incident powers $P_1 = 0$~dBm ($S = S^\text{L}$) and $P_2 = 12$~dBm.

\begin{figure*}
\centering
\includegraphics[width=2.05\columnwidth]{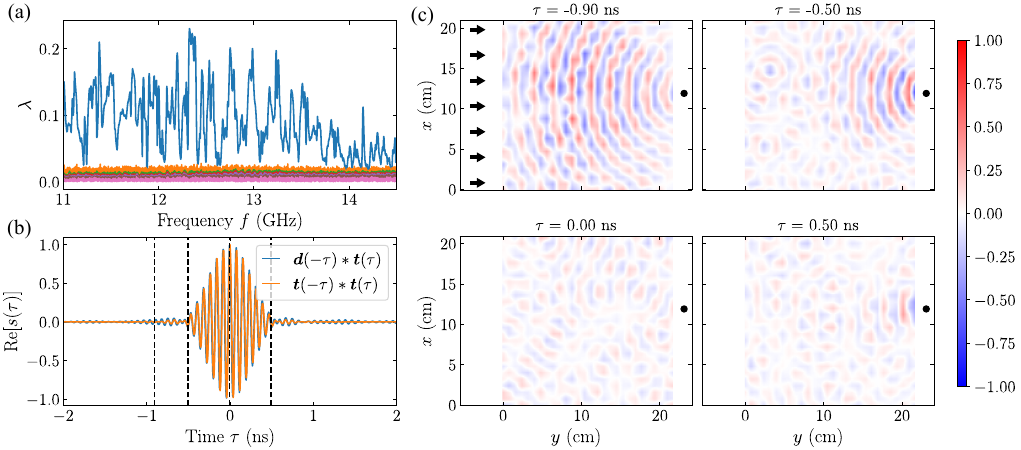}
\caption{\label{fig: exmpt spatiotemporal} Experimental results for a powered LNA connected to a single mode coax-to-waveguide transition located at the right interface of the cavity. (a) Spectrum of {the }singular values $\lambda_n$ {of $\Delta S$}. (b) Temporal signal at the nonlinear target corresponding to the back-propagation of the reconstructed signal for maximal focusing in space and time (blue line) and to a time-reversal experiment (orange line). {The maximum value of the cross-correlation between the two signals is 0.99}. (c) Real parts of the field at four times indicated by black dashed lines in (b) obtained from back-propagating {the first left singular vector }$\vec{d}^*$. The field around the position of the target (black dot) could not be probed since the port is located outside the cavity. The input antennas (indicated by black arrows) are located 30 cm toward the left of the map.
}
\end{figure*}

Two peaks in the spectrum of the first singular value $\lambda_1(\omega)$ of $\Delta S$ are observed at 12.68 and 13.35 GHz in Fig.~\ref{fig: exmpt 2D 1 target}(c), corresponding to resonances of the high-Q dielectric cylinder. The enhancement of the field intensity within the cylinder at resonance results in a strong coupling with the LNA, and thereby increases nonlinear effects.
We then scan the normalized field $t_n(x,y;\omega)$ inside the medium for each source channel $n$ by translating a short wire antenna in holes drilled into the top plate of the cavity. These measurements allow us to reconstruct the intensity map $\mathcal{I}(x,y;\omega)$ for any arbitrary incident wavefront $\vec{c}^\text{in}$, $\mathcal{I}(x,y;\omega) = | \vec{c}^{\text{in}*}(\omega) \cdot \vec{t}(x,y;\omega) |^2$. For the phase-conjugate of the first singular vector $\vec{c}^\text{in} = \vec{d}^*$, a strong enhancement of the intensity at the resonance $\omega_n$ is observed in Fig.~\ref{fig: exmpt 2D 1 target}(f) compared to random illumination in Fig.~\ref{fig: exmpt 2D 1 target}(e), since $\vec{d}$ gives the vector of Green's functions between the sources and the target. The intensity at the focus at 12.68 GHz is enhanced on average by a factor $\eta \simeq 4.7$ relative to a random incident wavefront.

Whereas we probed the system at two incident powers in order to apply a focus onto a target, we can instead exploit the reciprocity of linear systems for the detection of the nonlinearity.
For linear systems, the scattering matrix is symmetric; however, the nonreciprocity induced by the nonlinearity breaks the symmetry of $S^\alpha$~\cite{sounasFundamentalBoundsOperation2018,cotrufoNonlinearityinducedNonreciprocityPart2021,wangLossInducedViolationFundamental2023}, i.e.
\begin{equation}
    (\matrix{S}^\alpha - (\matrix{S}^\alpha)^T)_{m,n} = d_m  \alpha^{-1}f(\alpha d_n) - d_n \alpha^{-1} f(\alpha d_m).
\end{equation}
Thus, detecting the presence of a nonlinearity in practice requires a single measurement of $S(\omega)$ at high power by measuring the norm of the asymmetric part
\begin{equation}\label{eq:asymm_factor}
    \mathcal{A} = \norm{S^\alpha-(S^\alpha)^T}_F,
\end{equation}
where $\norm{\cdot}_F$ represents the Frobenius norm. At low power, $\mathcal{A}(\omega)$ is dominated by the noise level as the system is operating in the linear regime (see Fig.~\ref{fig: exmpt 2D 1 target}(d)). However, at high power, $\mathcal{A}(\omega)$ exhibits similar resonances as $\lambda_1(\omega)$. Note that this reciprocity condition is not completely able to extract $\vec{d}$. 
%More specifically, an SVD of the rank-two matrix $S^\alpha-(S^\alpha)^T$ provides us with the vector space spanned by $\vec{d}$ and $[f(\alpha d_n)]_n$. However, it is in general not possible to separate these states from each other without further measurements. Furthermore, $\mathcal{A}$ cannot give an indication on the number of nonlinearities within the system, only that nonlinear interactions have taken place.

We now demonstrate spatio-temporal focusing on a nonlinear target. The absorbing foams shown in Fig.~\ref{fig: exmpt 2D 1 target} are removed and the LNA is connected (without the resonator) to a single-mode waveguide located at the right interface. Apart from the nonlinear target, the cavity is now closed as the antennas at the right interface are terminated by open circuits (making them reflectors).
The LNA is now powered so that the {vector of} transmission coefficients $\vec{t}(\omega)$ to the target can also be measured by connecting the LNA to the eighth channel of the VNA. This is done only for the purpose of comparison as the measurement of $\vec{t}(\omega)$ is not necessary to determine the incident waveform. The nonlinearity is nonresonant as the antennas are matched over a broad frequency range (see Fig.~\ref{fig: exmpt spatiotemporal}(a)). Fluctuations in $\lambda_1(\omega)$ arise from multiple scattering within the cavity. 

Although the spatial wavefront for optimal focusing $\vec{c}^\text{in}(\omega) = \vec{d}^*(\omega)$ is obtained at each frequency through an SVD, {the frequency-dependence of the} global phase $\phi(\omega)$ {necessary} for optimal temporal focusing is still unknown.
We determine $\phi(\omega)$ using the procedure described in Ref.~\cite{yeoTimeReversalCommunications2022} that aligns the phase of each frequency component at the focal point (see Supplemental Material \cite{supp_mat}).  
The temporal signals found from the inverse Fourier transform 
$s_{\mathrm{opt}}(\tau)={\mathrm{FT}^{-1}[\vec{c}^\text{opt}(\omega) \cdot \vec{t}(\omega)]}$ 
%$ \red{S_{\mathrm{opt}}(t)=\psi_{\mathrm{opt}}^T(-t) \ast T(t)}$ 
and $s(\tau)=\mathrm{FT}^{-1}[\vec{t}^*(\omega) \cdot \vec{t}(\omega)]$ are in excellent agreement with each other, as shown in Fig.~\ref{fig: exmpt spatiotemporal}(b), demonstrating maximal focusing both in time and space on the nonlinear target. 
The back-propagated signals within the cavity $\mathcal{I}(x,y;\tau) = \mathrm{FT}^{-1}[\vec{c}^{\text{in}*} \cdot \vec{t}(x,y;\omega) ]$ are presented in Fig.~\ref{fig: exmpt spatiotemporal}(c). Interestingly, the amplitude of the outgoing field at positive times is strongly reduced relative to the incident field at negative times in Fig.~\ref{fig: exmpt spatiotemporal}(c-d), which indicates strong absorption within the lossy nonlinear target.

\begin{figure}
\centering
\includegraphics[width=.99\columnwidth]{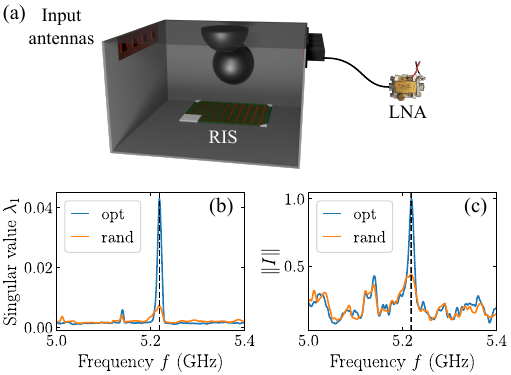}
\caption{\label{fig: exmpt 3D} (a) Schematic of the experiment. We measure the  scattering matrix $S$ of a chaotic cavity  using $N=7$ antennas.
(b) Optimization result for {the first singular value} $\lambda_1(f)$ {of $\Delta S$ (blue line)} compared with an average of 100 random configurations of the Reconfigurable Intelligent Surface (RIS) {(orange line)}.
%The inset shows the results for the asymmetry $\mathcal{A}(f)$.
(c) Same as (b) for the intensity measured at the target's position.}
\end{figure}

%In open environments for which the energy density illuminating a target is smaller than in confined geometries, wavefront shaping techniques may not be sufficient to detect the nonlinear signal. We thus investigate how the environment can be tuned to enhance the signal at the target~\cite{delhougneSpatiotemporalWaveFront2016,chenPerfectAbsorptionComplex2020,delhougneCoherentWaveControl2021}.
In environments for which the energy density illuminating a target is small, wavefront shaping techniques may not be sufficient to detect the nonlinear signal. We thus investigate how the environment can be tuned to enhance the signal at the target~\cite{delhougneSpatiotemporalWaveFront2016,chenPerfectAbsorptionComplex2020,delhougneCoherentWaveControl2021}.
For this purpose, we study a three-dimensional enclosure made programmable using a reconfigurable intelligent surface (RIS) (see Fig.~\ref{fig: exmpt 3D}(a)). 
For each of the 304 meta-atoms of the RIS, two states with a phase difference of roughly $\pi$ in reflection can be configured electronically. Seven antennas are used to measure a $7\times 7$ scattering matrix in the spectral window between 4.8 GHz and 5.8 GHz. The eighth antenna is connected to the nonlinear powered LNA. 

The metasurface is first optimized iteratively to maximize the first singular value $\lambda_1 (f)$ of $\Delta S$ at a single frequency $f_{\text{opt}} = $ 5.22 GHz. This corresponds to a modification of the Green's function inside the system such that for the wavefront {$\vec{d}$} the intensity at the position of the nonlinear device is increased. The result is shown in Fig.~\ref{fig: exmpt 3D} (b).
{At $f_{\text{opt}}$, the intensity enhancement is 2.3 fold compared to the average value for random configurations(Fig.~\ref{fig: exmpt 3D} (c)).}
%The optimization process also results in a strong enhancement of the asymmetry factor $\mathcal{A} (f_{\text{opt}})$ by a factor 3.5, and of the intensity at the target's position by a factor 2.3 (Fig.~\ref{fig: exmpt 3D} (c)).

\noindent \textit{Conclusion.---}
We have demonstrated that nonlinear elements embedded in a complex medium can be detected and localized using measurements of the scattering matrix at a single frequency for two incident powers (no harmonic generation is required). 
{The monochromatic aspect of this technique enables to detect any type of nonlinearity, and is therefore particularly relevant in cases where the specific nonlinear response  is unknown ($n$-th order harmonic generation, Kerr-like, etc.).}
We have shown that this noninvasive approach enables spatio-temporal focusing on nonlinear targets. 
{The experiments presented here were limited to confined geometries but our technique is directly applicable to open systems.}
Since smartphones contain nonlinear elements, this could provide a way to enhance the focusing of Wi-Fi signals on these devices in complex environments and may inspire new detection and localization setups.
{Our approach is broadly applicable to any kind of wave and can readily be used in any domain in which waves are used to probe a medium (such as acoustics, optics, ...). In particular, it provides an efficient way to obtain the focusing wavefront for wireless power transfer of bioelectronic devices~\cite{hoWirelessPowerTransfer2014}.}

%TC:ignore
\begin{acknowledgments}
This work is supported in part by the European Union through a European Regional Development Fund (ERDF), by the Ministry of Higher Education and Research, CNRS, Brittany region, Conseils Départementaux d’Ille-et-Vilaine and Côtes d’Armor, Rennes Métropole, and Lannion Trégor Communauté, through the CPER Project CyMoCod, in part by the French ``Agence Nationale de la Recherche" (ANR) under Grant ANR-22-CPJ1-0070-01, and in part by the Direction Générale de l'Armement through the Creach Labs under the project TRsweep. M.D. acknowledges the Institut Universitaire de France. The metasurface prototypes were purchased from Greenerwave. The authors acknowledge P. E. Davy for the 3D rendering of the experimental setup in Fig.~\ref{fig: exmpt 3D}(a).\nocite{yaghjian_electric_1980,PhysRevA.107.023112,chongHiddenBlackCoherent2011}
\end{acknowledgments}

% Create the reference section using BibTeX:
\bibliography{WS_NL}

%apsrev4-2.bst 2019-01-14 (MD) hand-edited version of apsrev4-1.bst
%Control: key (0)
%Control: author (8) initials jnrlst
%Control: editor formatted (1) identically to author
%Control: production of article title (0) allowed
%Control: page (0) single
%Control: year (1) truncated
%Control: production of eprint (1) enabled
\begin{thebibliography}{52}%
\makeatletter
\providecommand \@ifxundefined [1]{%
 \@ifx{#1\undefined}
}%
\providecommand \@ifnum [1]{%
 \ifnum #1\expandafter \@firstoftwo
 \else \expandafter \@secondoftwo
 \fi
}%
\providecommand \@ifx [1]{%
 \ifx #1\expandafter \@firstoftwo
 \else \expandafter \@secondoftwo
 \fi
}%
\providecommand \natexlab [1]{#1}%
\providecommand \enquote  [1]{``#1''}%
\providecommand \bibnamefont  [1]{#1}%
\providecommand \bibfnamefont [1]{#1}%
\providecommand \citenamefont [1]{#1}%
\providecommand \href@noop [0]{\@secondoftwo}%
\providecommand \href [0]{\begingroup \@sanitize@url \@href}%
\providecommand \@href[1]{\@@startlink{#1}\@@href}%
\providecommand \@@href[1]{\endgroup#1\@@endlink}%
\providecommand \@sanitize@url [0]{\catcode `\\12\catcode `\$12\catcode `\&12\catcode `\#12\catcode `\^12\catcode `\_12\catcode `\%12\relax}%
\providecommand \@@startlink[1]{}%
\providecommand \@@endlink[0]{}%
\providecommand \url  [0]{\begingroup\@sanitize@url \@url }%
\providecommand \@url [1]{\endgroup\@href {#1}{\urlprefix }}%
\providecommand \urlprefix  [0]{URL }%
\providecommand \Eprint [0]{\href }%
\providecommand \doibase [0]{https://doi.org/}%
\providecommand \selectlanguage [0]{\@gobble}%
\providecommand \bibinfo  [0]{\@secondoftwo}%
\providecommand \bibfield  [0]{\@secondoftwo}%
\providecommand \translation [1]{[#1]}%
\providecommand \BibitemOpen [0]{}%
\providecommand \bibitemStop [0]{}%
\providecommand \bibitemNoStop [0]{.\EOS\space}%
\providecommand \EOS [0]{\spacefactor3000\relax}%
\providecommand \BibitemShut  [1]{\csname bibitem#1\endcsname}%
\let\auto@bib@innerbib\@empty
%</preamble>
\bibitem [{\citenamefont {Mosk}\ \emph {et~al.}(2012)\citenamefont {Mosk}, \citenamefont {Lagendijk}, \citenamefont {Lerosey},\ and\ \citenamefont {Fink}}]{moskControllingWavesSpace2012}%
  \BibitemOpen
  \bibfield  {author} {\bibinfo {author} {\bibfnamefont {A.~P.}\ \bibnamefont {Mosk}}, \bibinfo {author} {\bibfnamefont {A.}~\bibnamefont {Lagendijk}}, \bibinfo {author} {\bibfnamefont {G.}~\bibnamefont {Lerosey}},\ and\ \bibinfo {author} {\bibfnamefont {M.}~\bibnamefont {Fink}},\ }\bibfield  {title} {\bibinfo {title} {Controlling waves in space and time for imaging and focusing in complex media},\ }\href {https://doi.org/10.1038/nphoton.2012.88} {\bibfield  {journal} {\bibinfo  {journal} {Nature Photon}\ }\textbf {\bibinfo {volume} {6}},\ \bibinfo {pages} {283} (\bibinfo {year} {2012})}\BibitemShut {NoStop}%
\bibitem [{\citenamefont {Rotter}\ and\ \citenamefont {Gigan}(2017)}]{rotterLightFieldsComplex2017}%
  \BibitemOpen
  \bibfield  {author} {\bibinfo {author} {\bibfnamefont {S.}~\bibnamefont {Rotter}}\ and\ \bibinfo {author} {\bibfnamefont {S.}~\bibnamefont {Gigan}},\ }\bibfield  {title} {\bibinfo {title} {Light fields in complex media: {{Mesoscopic}} scattering meets wave control},\ }\href {https://doi.org/10.1103/RevModPhys.89.015005} {\bibfield  {journal} {\bibinfo  {journal} {Rev. Mod. Phys.}\ }\textbf {\bibinfo {volume} {89}},\ \bibinfo {pages} {015005} (\bibinfo {year} {2017})}\BibitemShut {NoStop}%
\bibitem [{\citenamefont {Cao}\ \emph {et~al.}(2022)\citenamefont {Cao}, \citenamefont {Mosk},\ and\ \citenamefont {Rotter}}]{caoShapingPropagationLight2022}%
  \BibitemOpen
  \bibfield  {author} {\bibinfo {author} {\bibfnamefont {H.}~\bibnamefont {Cao}}, \bibinfo {author} {\bibfnamefont {A.~P.}\ \bibnamefont {Mosk}},\ and\ \bibinfo {author} {\bibfnamefont {S.}~\bibnamefont {Rotter}},\ }\bibfield  {title} {\bibinfo {title} {Shaping the propagation of light in complex media},\ }\href {https://doi.org/10.1038/s41567-022-01677-x} {\bibfield  {journal} {\bibinfo  {journal} {Nat. Phys.}\ }\textbf {\bibinfo {volume} {18}},\ \bibinfo {pages} {994} (\bibinfo {year} {2022})}\BibitemShut {NoStop}%
\bibitem [{\citenamefont {Vellekoop}\ \emph {et~al.}(2010)\citenamefont {Vellekoop}, \citenamefont {Lagendijk},\ and\ \citenamefont {Mosk}}]{vellekoopExploitingDisorderPerfect2010}%
  \BibitemOpen
  \bibfield  {author} {\bibinfo {author} {\bibfnamefont {I.~M.}\ \bibnamefont {Vellekoop}}, \bibinfo {author} {\bibfnamefont {A.}~\bibnamefont {Lagendijk}},\ and\ \bibinfo {author} {\bibfnamefont {A.~P.}\ \bibnamefont {Mosk}},\ }\bibfield  {title} {\bibinfo {title} {Exploiting disorder for perfect focusing},\ }\href {https://doi.org/10.1038/nphoton.2010.3} {\bibfield  {journal} {\bibinfo  {journal} {Nature Photon}\ }\textbf {\bibinfo {volume} {4}},\ \bibinfo {pages} {320} (\bibinfo {year} {2010})}\BibitemShut {NoStop}%
\bibitem [{\citenamefont {Jeong}\ \emph {et~al.}(2018)\citenamefont {Jeong}, \citenamefont {Lee}, \citenamefont {Choi}, \citenamefont {Kang}, \citenamefont {Hong}, \citenamefont {Park}, \citenamefont {Lim}, \citenamefont {Park},\ and\ \citenamefont {Choi}}]{jeongFocusingLightEnergy2018}%
  \BibitemOpen
  \bibfield  {author} {\bibinfo {author} {\bibfnamefont {S.}~\bibnamefont {Jeong}}, \bibinfo {author} {\bibfnamefont {Y.-R.}\ \bibnamefont {Lee}}, \bibinfo {author} {\bibfnamefont {W.}~\bibnamefont {Choi}}, \bibinfo {author} {\bibfnamefont {S.}~\bibnamefont {Kang}}, \bibinfo {author} {\bibfnamefont {J.~H.}\ \bibnamefont {Hong}}, \bibinfo {author} {\bibfnamefont {J.-S.}\ \bibnamefont {Park}}, \bibinfo {author} {\bibfnamefont {Y.-S.}\ \bibnamefont {Lim}}, \bibinfo {author} {\bibfnamefont {H.-G.}\ \bibnamefont {Park}},\ and\ \bibinfo {author} {\bibfnamefont {W.}~\bibnamefont {Choi}},\ }\bibfield  {title} {\bibinfo {title} {Focusing of light energy inside a scattering medium by controlling the time-gated multiple light scattering},\ }\href {https://doi.org/10.1038/s41566-018-0120-9} {\bibfield  {journal} {\bibinfo  {journal} {Nature Photon}\ }\textbf {\bibinfo {volume} {12}},\ \bibinfo {pages} {277} (\bibinfo {year} {2018})}\BibitemShut {NoStop}%
\bibitem [{\citenamefont {Bender}\ \emph {et~al.}(2022)\citenamefont {Bender}, \citenamefont {Yamilov}, \citenamefont {Goetschy}, \citenamefont {Y{\i}lmaz}, \citenamefont {Hsu},\ and\ \citenamefont {Cao}}]{benderDepthtargetedEnergyDelivery2022}%
  \BibitemOpen
  \bibfield  {author} {\bibinfo {author} {\bibfnamefont {N.}~\bibnamefont {Bender}}, \bibinfo {author} {\bibfnamefont {A.}~\bibnamefont {Yamilov}}, \bibinfo {author} {\bibfnamefont {A.}~\bibnamefont {Goetschy}}, \bibinfo {author} {\bibfnamefont {H.}~\bibnamefont {Y{\i}lmaz}}, \bibinfo {author} {\bibfnamefont {C.~W.}\ \bibnamefont {Hsu}},\ and\ \bibinfo {author} {\bibfnamefont {H.}~\bibnamefont {Cao}},\ }\bibfield  {title} {\bibinfo {title} {Depth-targeted energy delivery deep inside scattering media},\ }\href {https://doi.org/10.1038/s41567-021-01475-x} {\bibfield  {journal} {\bibinfo  {journal} {Nat. Phys.}\ }\textbf {\bibinfo {volume} {18}},\ \bibinfo {pages} {309} (\bibinfo {year} {2022})}\BibitemShut {NoStop}%
\bibitem [{\citenamefont {Yaqoob}\ \emph {et~al.}(2008)\citenamefont {Yaqoob}, \citenamefont {Psaltis}, \citenamefont {Feld},\ and\ \citenamefont {Yang}}]{yaqoobOpticalPhaseConjugation2008}%
  \BibitemOpen
  \bibfield  {author} {\bibinfo {author} {\bibfnamefont {Z.}~\bibnamefont {Yaqoob}}, \bibinfo {author} {\bibfnamefont {D.}~\bibnamefont {Psaltis}}, \bibinfo {author} {\bibfnamefont {M.~S.}\ \bibnamefont {Feld}},\ and\ \bibinfo {author} {\bibfnamefont {C.}~\bibnamefont {Yang}},\ }\bibfield  {title} {\bibinfo {title} {Optical phase conjugation for turbidity suppression in biological samples},\ }\href {https://doi.org/10.1038/nphoton.2007.297} {\bibfield  {journal} {\bibinfo  {journal} {Nature Photon}\ }\textbf {\bibinfo {volume} {2}},\ \bibinfo {pages} {110} (\bibinfo {year} {2008})}\BibitemShut {NoStop}%
\bibitem [{\citenamefont {Popoff}\ \emph {et~al.}(2010)\citenamefont {Popoff}, \citenamefont {Lerosey}, \citenamefont {Carminati}, \citenamefont {Fink}, \citenamefont {Boccara},\ and\ \citenamefont {Gigan}}]{popoffMeasuringTransmissionMatrix2010}%
  \BibitemOpen
  \bibfield  {author} {\bibinfo {author} {\bibfnamefont {S.~M.}\ \bibnamefont {Popoff}}, \bibinfo {author} {\bibfnamefont {G.}~\bibnamefont {Lerosey}}, \bibinfo {author} {\bibfnamefont {R.}~\bibnamefont {Carminati}}, \bibinfo {author} {\bibfnamefont {M.}~\bibnamefont {Fink}}, \bibinfo {author} {\bibfnamefont {A.~C.}\ \bibnamefont {Boccara}},\ and\ \bibinfo {author} {\bibfnamefont {S.}~\bibnamefont {Gigan}},\ }\bibfield  {title} {\bibinfo {title} {Measuring the {{Transmission Matrix}} in {{Optics}}: {{An Approach}} to the {{Study}} and {{Control}} of {{Light Propagation}} in {{Disordered Media}}},\ }\href {https://doi.org/10.1103/PhysRevLett.104.100601} {\bibfield  {journal} {\bibinfo  {journal} {Phys. Rev. Lett.}\ }\textbf {\bibinfo {volume} {104}},\ \bibinfo {pages} {100601} (\bibinfo {year} {2010})}\BibitemShut {NoStop}%
\bibitem [{\citenamefont {Xu}\ \emph {et~al.}(2011)\citenamefont {Xu}, \citenamefont {Liu},\ and\ \citenamefont {Wang}}]{xuTimereversedUltrasonicallyEncoded2011}%
  \BibitemOpen
  \bibfield  {author} {\bibinfo {author} {\bibfnamefont {X.}~\bibnamefont {Xu}}, \bibinfo {author} {\bibfnamefont {H.}~\bibnamefont {Liu}},\ and\ \bibinfo {author} {\bibfnamefont {L.~V.}\ \bibnamefont {Wang}},\ }\bibfield  {title} {\bibinfo {title} {Time-reversed ultrasonically encoded optical focusing into scattering media},\ }\href {https://doi.org/10.1038/nphoton.2010.306} {\bibfield  {journal} {\bibinfo  {journal} {Nature Photon}\ }\textbf {\bibinfo {volume} {5}},\ \bibinfo {pages} {154} (\bibinfo {year} {2011})}\BibitemShut {NoStop}%
\bibitem [{\citenamefont {Hsu}\ \emph {et~al.}(2016)\citenamefont {Hsu}, \citenamefont {Liew}, \citenamefont {Goetschy}, \citenamefont {Cao},\ and\ \citenamefont {Stone}}]{hsuCorrelationenhancedControlWave2016}%
  \BibitemOpen
  \bibfield  {author} {\bibinfo {author} {\bibfnamefont {C.~W.}\ \bibnamefont {Hsu}}, \bibinfo {author} {\bibfnamefont {S.~F.}\ \bibnamefont {Liew}}, \bibinfo {author} {\bibfnamefont {A.}~\bibnamefont {Goetschy}}, \bibinfo {author} {\bibfnamefont {H.}~\bibnamefont {Cao}},\ and\ \bibinfo {author} {\bibfnamefont {A.~D.}\ \bibnamefont {Stone}},\ }\bibfield  {title} {\bibinfo {title} {Correlation-enhanced control of wave focusing in disordered media},\ }\bibfield  {journal} {\bibinfo  {journal} {Nat. Phys.}\ }\href {https://doi.org/10.1038/NPHYS4036} {10.1038/NPHYS4036} (\bibinfo {year} {2016})\BibitemShut {NoStop}%
\bibitem [{\citenamefont {Derode}\ \emph {et~al.}(1995)\citenamefont {Derode}, \citenamefont {Roux},\ and\ \citenamefont {Fink}}]{derodeRobustAcousticTime1995}%
  \BibitemOpen
  \bibfield  {author} {\bibinfo {author} {\bibfnamefont {A.}~\bibnamefont {Derode}}, \bibinfo {author} {\bibfnamefont {P.}~\bibnamefont {Roux}},\ and\ \bibinfo {author} {\bibfnamefont {M.}~\bibnamefont {Fink}},\ }\bibfield  {title} {\bibinfo {title} {Robust {{Acoustic Time Reversal}} with {{High-Order Multiple Scattering}}},\ }\href {https://doi.org/10.1103/PhysRevLett.75.4206} {\bibfield  {journal} {\bibinfo  {journal} {Phys. Rev. Lett.}\ }\textbf {\bibinfo {volume} {75}},\ \bibinfo {pages} {4206} (\bibinfo {year} {1995})}\BibitemShut {NoStop}%
\bibitem [{\citenamefont {Kim}\ \emph {et~al.}(2012)\citenamefont {Kim}, \citenamefont {Choi}, \citenamefont {Yoon}, \citenamefont {Choi}, \citenamefont {Kim}, \citenamefont {Park},\ and\ \citenamefont {Choi}}]{kimMaximalEnergyTransport2012}%
  \BibitemOpen
  \bibfield  {author} {\bibinfo {author} {\bibfnamefont {M.}~\bibnamefont {Kim}}, \bibinfo {author} {\bibfnamefont {Y.}~\bibnamefont {Choi}}, \bibinfo {author} {\bibfnamefont {C.}~\bibnamefont {Yoon}}, \bibinfo {author} {\bibfnamefont {W.}~\bibnamefont {Choi}}, \bibinfo {author} {\bibfnamefont {J.}~\bibnamefont {Kim}}, \bibinfo {author} {\bibfnamefont {Q.-H.}\ \bibnamefont {Park}},\ and\ \bibinfo {author} {\bibfnamefont {W.}~\bibnamefont {Choi}},\ }\bibfield  {title} {\bibinfo {title} {Maximal energy transport through disordered media with the implementation of transmission eigenchannels},\ }\href {https://doi.org/10.1038/nphoton.2012.159} {\bibfield  {journal} {\bibinfo  {journal} {Nature Photon}\ }\textbf {\bibinfo {volume} {6}},\ \bibinfo {pages} {581} (\bibinfo {year} {2012})}\BibitemShut {NoStop}%
\bibitem [{\citenamefont {Lambert}\ \emph {et~al.}(2020)\citenamefont {Lambert}, \citenamefont {Cobus}, \citenamefont {Frappart}, \citenamefont {Fink},\ and\ \citenamefont {Aubry}}]{lambertDistortionMatrixApproach2020}%
  \BibitemOpen
  \bibfield  {author} {\bibinfo {author} {\bibfnamefont {W.}~\bibnamefont {Lambert}}, \bibinfo {author} {\bibfnamefont {L.~A.}\ \bibnamefont {Cobus}}, \bibinfo {author} {\bibfnamefont {T.}~\bibnamefont {Frappart}}, \bibinfo {author} {\bibfnamefont {M.}~\bibnamefont {Fink}},\ and\ \bibinfo {author} {\bibfnamefont {A.}~\bibnamefont {Aubry}},\ }\bibfield  {title} {\bibinfo {title} {Distortion matrix approach for ultrasound imaging of random scattering media},\ }\href {https://doi.org/10.1073/pnas.1921533117} {\bibfield  {journal} {\bibinfo  {journal} {Proc. Natl. Acad. Sci.}\ }\textbf {\bibinfo {volume} {117}},\ \bibinfo {pages} {14645} (\bibinfo {year} {2020})}\BibitemShut {NoStop}%
\bibitem [{\citenamefont {Judkewitz}\ \emph {et~al.}(2013)\citenamefont {Judkewitz}, \citenamefont {Wang}, \citenamefont {Horstmeyer}, \citenamefont {Mathy},\ and\ \citenamefont {Yang}}]{judkewitzSpecklescaleFocusingDiffusive2013}%
  \BibitemOpen
  \bibfield  {author} {\bibinfo {author} {\bibfnamefont {B.}~\bibnamefont {Judkewitz}}, \bibinfo {author} {\bibfnamefont {Y.~M.}\ \bibnamefont {Wang}}, \bibinfo {author} {\bibfnamefont {R.}~\bibnamefont {Horstmeyer}}, \bibinfo {author} {\bibfnamefont {A.}~\bibnamefont {Mathy}},\ and\ \bibinfo {author} {\bibfnamefont {C.}~\bibnamefont {Yang}},\ }\bibfield  {title} {\bibinfo {title} {Speckle-scale focusing in the diffusive regime with time reversal of variance-encoded light ({{TROVE}})},\ }\href {https://doi.org/10.1038/nphoton.2013.31} {\bibfield  {journal} {\bibinfo  {journal} {Nature Photon}\ }\textbf {\bibinfo {volume} {7}},\ \bibinfo {pages} {300} (\bibinfo {year} {2013})}\BibitemShut {NoStop}%
\bibitem [{\citenamefont {Lai}\ \emph {et~al.}(2015)\citenamefont {Lai}, \citenamefont {Wang}, \citenamefont {Tay},\ and\ \citenamefont {Wang}}]{laiPhotoacousticallyGuidedWavefront2015}%
  \BibitemOpen
  \bibfield  {author} {\bibinfo {author} {\bibfnamefont {P.}~\bibnamefont {Lai}}, \bibinfo {author} {\bibfnamefont {L.}~\bibnamefont {Wang}}, \bibinfo {author} {\bibfnamefont {J.~W.}\ \bibnamefont {Tay}},\ and\ \bibinfo {author} {\bibfnamefont {L.~V.}\ \bibnamefont {Wang}},\ }\bibfield  {title} {\bibinfo {title} {Photoacoustically guided wavefront shaping for enhanced optical focusing in scattering media},\ }\href {https://doi.org/10.1038/nphoton.2014.322} {\bibfield  {journal} {\bibinfo  {journal} {Nature Photon}\ }\textbf {\bibinfo {volume} {9}},\ \bibinfo {pages} {126} (\bibinfo {year} {2015})}\BibitemShut {NoStop}%
\bibitem [{\citenamefont {Horstmeyer}\ \emph {et~al.}(2015)\citenamefont {Horstmeyer}, \citenamefont {Ruan},\ and\ \citenamefont {Yang}}]{horstmeyerGuidestarassistedWavefrontshapingMethods2015}%
  \BibitemOpen
  \bibfield  {author} {\bibinfo {author} {\bibfnamefont {R.}~\bibnamefont {Horstmeyer}}, \bibinfo {author} {\bibfnamefont {H.}~\bibnamefont {Ruan}},\ and\ \bibinfo {author} {\bibfnamefont {C.}~\bibnamefont {Yang}},\ }\bibfield  {title} {\bibinfo {title} {Guidestar-assisted wavefront-shaping methods for focusing light into biological tissue},\ }\href {https://doi.org/10.1038/nphoton.2015.140} {\bibfield  {journal} {\bibinfo  {journal} {Nature Photon}\ }\textbf {\bibinfo {volume} {9}},\ \bibinfo {pages} {563} (\bibinfo {year} {2015})}\BibitemShut {NoStop}%
\bibitem [{\citenamefont {Miller}(1981)}]{millerUltrasonicDetectionResonant1981}%
  \BibitemOpen
  \bibfield  {author} {\bibinfo {author} {\bibfnamefont {D.}~\bibnamefont {Miller}},\ }\bibfield  {title} {\bibinfo {title} {Ultrasonic detection of resonant cavitation bubbles in a flow tube by their second-harmonic emissions},\ }\href {https://doi.org/10.1016/0041-624X(81)90006-8} {\bibfield  {journal} {\bibinfo  {journal} {Ultrasonics}\ }\textbf {\bibinfo {volume} {19}},\ \bibinfo {pages} {217} (\bibinfo {year} {1981})}\BibitemShut {NoStop}%
\bibitem [{\citenamefont {Qin}\ \emph {et~al.}(2009)\citenamefont {Qin}, \citenamefont {Caskey},\ and\ \citenamefont {Ferrara}}]{qinUltrasoundContrastMicrobubbles2009}%
  \BibitemOpen
  \bibfield  {author} {\bibinfo {author} {\bibfnamefont {S.}~\bibnamefont {Qin}}, \bibinfo {author} {\bibfnamefont {C.~F.}\ \bibnamefont {Caskey}},\ and\ \bibinfo {author} {\bibfnamefont {K.~W.}\ \bibnamefont {Ferrara}},\ }\bibfield  {title} {\bibinfo {title} {Ultrasound contrast microbubbles in imaging and therapy: Physical principles and engineering},\ }\href {https://doi.org/10.1088/0031-9155/54/6/R01} {\bibfield  {journal} {\bibinfo  {journal} {Phys. Med. Biol.}\ }\textbf {\bibinfo {volume} {54}},\ \bibinfo {pages} {R27} (\bibinfo {year} {2009})}\BibitemShut {NoStop}%
\bibitem [{\citenamefont {Evans}\ and\ \citenamefont {Xie}(2008)}]{evansCoherentAntiStokesRaman2008}%
  \BibitemOpen
  \bibfield  {author} {\bibinfo {author} {\bibfnamefont {C.~L.}\ \bibnamefont {Evans}}\ and\ \bibinfo {author} {\bibfnamefont {X.~S.}\ \bibnamefont {Xie}},\ }\bibfield  {title} {\bibinfo {title} {Coherent {{Anti-Stokes Raman Scattering Microscopy}}: {{Chemical Imaging}} for {{Biology}} and {{Medicine}}},\ }\href {https://doi.org/10.1146/annurev.anchem.1.031207.112754} {\bibfield  {journal} {\bibinfo  {journal} {Annual Rev. Anal. Chem.}\ }\textbf {\bibinfo {volume} {1}},\ \bibinfo {pages} {883} (\bibinfo {year} {2008})}\BibitemShut {NoStop}%
\bibitem [{\citenamefont {Katz}\ \emph {et~al.}(2014)\citenamefont {Katz}, \citenamefont {Small}, \citenamefont {Guan},\ and\ \citenamefont {Silberberg}}]{katzNoninvasiveNonlinearFocusing2014}%
  \BibitemOpen
  \bibfield  {author} {\bibinfo {author} {\bibfnamefont {O.}~\bibnamefont {Katz}}, \bibinfo {author} {\bibfnamefont {E.}~\bibnamefont {Small}}, \bibinfo {author} {\bibfnamefont {Y.}~\bibnamefont {Guan}},\ and\ \bibinfo {author} {\bibfnamefont {Y.}~\bibnamefont {Silberberg}},\ }\bibfield  {title} {\bibinfo {title} {Noninvasive nonlinear focusing and imaging through strongly scattering turbid layers},\ }\href {https://doi.org/10.1364/OPTICA.1.000170} {\bibfield  {journal} {\bibinfo  {journal} {Optica}\ }\textbf {\bibinfo {volume} {1}},\ \bibinfo {pages} {170} (\bibinfo {year} {2014})}\BibitemShut {NoStop}%
\bibitem [{\citenamefont {De~Aguiar}\ \emph {et~al.}(2016)\citenamefont {De~Aguiar}, \citenamefont {Brasselet},\ and\ \citenamefont {Gigan}}]{deaguiarEnhancedNonlinearImaging2016}%
  \BibitemOpen
  \bibfield  {author} {\bibinfo {author} {\bibfnamefont {H.~B.}\ \bibnamefont {De~Aguiar}}, \bibinfo {author} {\bibfnamefont {S.}~\bibnamefont {Brasselet}},\ and\ \bibinfo {author} {\bibfnamefont {S.}~\bibnamefont {Gigan}},\ }\bibfield  {title} {\bibinfo {title} {Enhanced nonlinear imaging through scattering media using transmission-matrix-based wave-front shaping},\ }\href {https://doi.org/10.1103/PhysRevA.94.043830} {\bibfield  {journal} {\bibinfo  {journal} {Phys. Rev. A}\ }\textbf {\bibinfo {volume} {94}},\ \bibinfo {pages} {043830} (\bibinfo {year} {2016})}\BibitemShut {NoStop}%
\bibitem [{\citenamefont {Moon}\ \emph {et~al.}(2023)\citenamefont {Moon}, \citenamefont {Cho}, \citenamefont {Kang}, \citenamefont {Jang},\ and\ \citenamefont {Choi}}]{moonMeasuringScatteringTensor2023}%
  \BibitemOpen
  \bibfield  {author} {\bibinfo {author} {\bibfnamefont {J.}~\bibnamefont {Moon}}, \bibinfo {author} {\bibfnamefont {Y.-C.}\ \bibnamefont {Cho}}, \bibinfo {author} {\bibfnamefont {S.}~\bibnamefont {Kang}}, \bibinfo {author} {\bibfnamefont {M.}~\bibnamefont {Jang}},\ and\ \bibinfo {author} {\bibfnamefont {W.}~\bibnamefont {Choi}},\ }\bibfield  {title} {\bibinfo {title} {Measuring the scattering tensor of a disordered nonlinear medium},\ }\href {https://doi.org/10.1038/s41567-023-02163-8} {\bibfield  {journal} {\bibinfo  {journal} {Nat. Phys.}\ }\textbf {\bibinfo {volume} {19}},\ \bibinfo {pages} {1709} (\bibinfo {year} {2023})}\BibitemShut {NoStop}%
\bibitem [{\citenamefont {Yoon}\ \emph {et~al.}(2020)\citenamefont {Yoon}, \citenamefont {Kim}, \citenamefont {Jang}, \citenamefont {Choi}, \citenamefont {Choi}, \citenamefont {Kang},\ and\ \citenamefont {Choi}}]{yoonDeepOpticalImaging2020}%
  \BibitemOpen
  \bibfield  {author} {\bibinfo {author} {\bibfnamefont {S.}~\bibnamefont {Yoon}}, \bibinfo {author} {\bibfnamefont {M.}~\bibnamefont {Kim}}, \bibinfo {author} {\bibfnamefont {M.}~\bibnamefont {Jang}}, \bibinfo {author} {\bibfnamefont {Y.}~\bibnamefont {Choi}}, \bibinfo {author} {\bibfnamefont {W.}~\bibnamefont {Choi}}, \bibinfo {author} {\bibfnamefont {S.}~\bibnamefont {Kang}},\ and\ \bibinfo {author} {\bibfnamefont {W.}~\bibnamefont {Choi}},\ }\bibfield  {title} {\bibinfo {title} {Deep optical imaging within complex scattering media},\ }\href {https://doi.org/10.1038/s42254-019-0143-2} {\bibfield  {journal} {\bibinfo  {journal} {Nat Rev Phys}\ }\textbf {\bibinfo {volume} {2}},\ \bibinfo {pages} {141} (\bibinfo {year} {2020})}\BibitemShut {NoStop}%
\bibitem [{\citenamefont {Reisner}\ \emph {et~al.}(2020)\citenamefont {Reisner}, \citenamefont {Jeon}, \citenamefont {Schindler}, \citenamefont {Schomerus}, \citenamefont {Mortessagne}, \citenamefont {Kuhl},\ and\ \citenamefont {Kottos}}]{reisnerSelfshieldedTopologicalReceiver2020}%
  \BibitemOpen
  \bibfield  {author} {\bibinfo {author} {\bibfnamefont {M.}~\bibnamefont {Reisner}}, \bibinfo {author} {\bibfnamefont {D.~H.}\ \bibnamefont {Jeon}}, \bibinfo {author} {\bibfnamefont {C.}~\bibnamefont {Schindler}}, \bibinfo {author} {\bibfnamefont {H.}~\bibnamefont {Schomerus}}, \bibinfo {author} {\bibfnamefont {F.}~\bibnamefont {Mortessagne}}, \bibinfo {author} {\bibfnamefont {U.}~\bibnamefont {Kuhl}},\ and\ \bibinfo {author} {\bibfnamefont {T.}~\bibnamefont {Kottos}},\ }\bibfield  {title} {\bibinfo {title} {Self-{{Shielded Topological Receiver Protectors}}},\ }\href {https://doi.org/10.1103/PhysRevApplied.13.034067} {\bibfield  {journal} {\bibinfo  {journal} {Phys. Rev. Applied}\ }\textbf {\bibinfo {volume} {13}},\ \bibinfo {pages} {034067} (\bibinfo {year} {2020})}\BibitemShut {NoStop}%
\bibitem [{\citenamefont {Jeon}\ \emph {et~al.}(2020)\citenamefont {Jeon}, \citenamefont {Reisner}, \citenamefont {Mortessagne}, \citenamefont {Kottos},\ and\ \citenamefont {Kuhl}}]{jeonNonHermitianSymmetricSpectral2020}%
  \BibitemOpen
  \bibfield  {author} {\bibinfo {author} {\bibfnamefont {D.~H.}\ \bibnamefont {Jeon}}, \bibinfo {author} {\bibfnamefont {M.}~\bibnamefont {Reisner}}, \bibinfo {author} {\bibfnamefont {F.}~\bibnamefont {Mortessagne}}, \bibinfo {author} {\bibfnamefont {T.}~\bibnamefont {Kottos}},\ and\ \bibinfo {author} {\bibfnamefont {U.}~\bibnamefont {Kuhl}},\ }\bibfield  {title} {\bibinfo {title} {Non-{{Hermitian C T}} -{{Symmetric Spectral Protection}} of {{Nonlinear Defect Modes}}},\ }\href {https://doi.org/10.1103/PhysRevLett.125.113901} {\bibfield  {journal} {\bibinfo  {journal} {Phys. Rev. Lett.}\ }\textbf {\bibinfo {volume} {125}},\ \bibinfo {pages} {113901} (\bibinfo {year} {2020})}\BibitemShut {NoStop}%
\bibitem [{\citenamefont {Suwunnarat}\ \emph {et~al.}(2022)\citenamefont {Suwunnarat}, \citenamefont {Tang}, \citenamefont {Reisner}, \citenamefont {Mortessagne}, \citenamefont {Kuhl},\ and\ \citenamefont {Kottos}}]{suwunnaratNonlinearCoherentPerfect2022}%
  \BibitemOpen
  \bibfield  {author} {\bibinfo {author} {\bibfnamefont {S.}~\bibnamefont {Suwunnarat}}, \bibinfo {author} {\bibfnamefont {Y.}~\bibnamefont {Tang}}, \bibinfo {author} {\bibfnamefont {M.}~\bibnamefont {Reisner}}, \bibinfo {author} {\bibfnamefont {F.}~\bibnamefont {Mortessagne}}, \bibinfo {author} {\bibfnamefont {U.}~\bibnamefont {Kuhl}},\ and\ \bibinfo {author} {\bibfnamefont {T.}~\bibnamefont {Kottos}},\ }\bibfield  {title} {\bibinfo {title} {Non-linear coherent perfect absorption in the proximity of exceptional points},\ }\href {https://doi.org/10.1038/s42005-021-00782-2} {\bibfield  {journal} {\bibinfo  {journal} {Commun Phys}\ }\textbf {\bibinfo {volume} {5}},\ \bibinfo {pages} {5} (\bibinfo {year} {2022})}\BibitemShut {NoStop}%
\bibitem [{\citenamefont {Mazzaro}\ \emph {et~al.}(2017)\citenamefont {Mazzaro}, \citenamefont {Martone}, \citenamefont {Ranney},\ and\ \citenamefont {Narayanan}}]{mazzaroNonlinearRadarFinding2017}%
  \BibitemOpen
  \bibfield  {author} {\bibinfo {author} {\bibfnamefont {G.~J.}\ \bibnamefont {Mazzaro}}, \bibinfo {author} {\bibfnamefont {A.~F.}\ \bibnamefont {Martone}}, \bibinfo {author} {\bibfnamefont {K.~I.}\ \bibnamefont {Ranney}},\ and\ \bibinfo {author} {\bibfnamefont {R.~M.}\ \bibnamefont {Narayanan}},\ }\bibfield  {title} {\bibinfo {title} {Nonlinear {{Radar}} for {{Finding RF Electronics}}: {{System Design}} and {{Recent Advancements}}},\ }\href {https://doi.org/10.1109/TMTT.2016.2640953} {\bibfield  {journal} {\bibinfo  {journal} {IEEE Trans. Microwave Theory Techn.}\ }\textbf {\bibinfo {volume} {65}},\ \bibinfo {pages} {1716} (\bibinfo {year} {2017})}\BibitemShut {NoStop}%
\bibitem [{\citenamefont {Perez}\ \emph {et~al.}(2022)\citenamefont {Perez}, \citenamefont {Mazzaro}, \citenamefont {Pierson},\ and\ \citenamefont {Kotz}}]{perezDetectingPresenceElectronic2022}%
  \BibitemOpen
  \bibfield  {author} {\bibinfo {author} {\bibfnamefont {B.}~\bibnamefont {Perez}}, \bibinfo {author} {\bibfnamefont {G.}~\bibnamefont {Mazzaro}}, \bibinfo {author} {\bibfnamefont {T.~J.}\ \bibnamefont {Pierson}},\ and\ \bibinfo {author} {\bibfnamefont {D.}~\bibnamefont {Kotz}},\ }\bibfield  {title} {\bibinfo {title} {Detecting the {{Presence}} of {{Electronic Devices}} in {{Smart Homes Using Harmonic Radar Technology}}},\ }\href {https://doi.org/10.3390/rs14020327} {\bibfield  {journal} {\bibinfo  {journal} {Remote Sensing}\ }\textbf {\bibinfo {volume} {14}},\ \bibinfo {pages} {327} (\bibinfo {year} {2022})}\BibitemShut {NoStop}%
\bibitem [{\citenamefont {Ambichl}\ \emph {et~al.}(2017)\citenamefont {Ambichl}, \citenamefont {Brandst{\"o}tter}, \citenamefont {B{\"o}hm}, \citenamefont {K{\"u}hmayer}, \citenamefont {Kuhl},\ and\ \citenamefont {Rotter}}]{ambichlFocusingDisorderedMedia2017}%
  \BibitemOpen
  \bibfield  {author} {\bibinfo {author} {\bibfnamefont {P.}~\bibnamefont {Ambichl}}, \bibinfo {author} {\bibfnamefont {A.}~\bibnamefont {Brandst{\"o}tter}}, \bibinfo {author} {\bibfnamefont {J.}~\bibnamefont {B{\"o}hm}}, \bibinfo {author} {\bibfnamefont {M.}~\bibnamefont {K{\"u}hmayer}}, \bibinfo {author} {\bibfnamefont {U.}~\bibnamefont {Kuhl}},\ and\ \bibinfo {author} {\bibfnamefont {S.}~\bibnamefont {Rotter}},\ }\bibfield  {title} {\bibinfo {title} {Focusing inside {{Disordered Media}} with the {{Generalized Wigner-Smith Operator}}},\ }\href {https://doi.org/10.1103/PhysRevLett.119.033903} {\bibfield  {journal} {\bibinfo  {journal} {Phys. Rev. Lett.}\ }\textbf {\bibinfo {volume} {119}},\ \bibinfo {pages} {033903} (\bibinfo {year} {2017})}\BibitemShut {NoStop}%
\bibitem [{\citenamefont {Horodynski}\ \emph {et~al.}(2020)\citenamefont {Horodynski}, \citenamefont {K{\"u}hmayer}, \citenamefont {Brandst{\"o}tter}, \citenamefont {Pichler}, \citenamefont {Fyodorov}, \citenamefont {Kuhl},\ and\ \citenamefont {Rotter}}]{horodynskiOptimalWaveFields2020}%
  \BibitemOpen
  \bibfield  {author} {\bibinfo {author} {\bibfnamefont {M.}~\bibnamefont {Horodynski}}, \bibinfo {author} {\bibfnamefont {M.}~\bibnamefont {K{\"u}hmayer}}, \bibinfo {author} {\bibfnamefont {A.}~\bibnamefont {Brandst{\"o}tter}}, \bibinfo {author} {\bibfnamefont {K.}~\bibnamefont {Pichler}}, \bibinfo {author} {\bibfnamefont {Y.~V.}\ \bibnamefont {Fyodorov}}, \bibinfo {author} {\bibfnamefont {U.}~\bibnamefont {Kuhl}},\ and\ \bibinfo {author} {\bibfnamefont {S.}~\bibnamefont {Rotter}},\ }\bibfield  {title} {\bibinfo {title} {Optimal wave fields for micromanipulation in complex scattering environments},\ }\href {https://doi.org/10.1038/s41566-019-0550-z} {\bibfield  {journal} {\bibinfo  {journal} {Nat. Photonics}\ }\textbf {\bibinfo {volume} {14}},\ \bibinfo {pages} {149} (\bibinfo {year} {2020})}\BibitemShut {NoStop}%
\bibitem [{\citenamefont {Del~Hougne}\ \emph {et~al.}(2021{\natexlab{a}})\citenamefont {Del~Hougne}, \citenamefont {Yeo}, \citenamefont {Besnier},\ and\ \citenamefont {Davy}}]{delhougneCoherentWaveControl2021}%
  \BibitemOpen
  \bibfield  {author} {\bibinfo {author} {\bibfnamefont {P.}~\bibnamefont {Del~Hougne}}, \bibinfo {author} {\bibfnamefont {K.~B.}\ \bibnamefont {Yeo}}, \bibinfo {author} {\bibfnamefont {P.}~\bibnamefont {Besnier}},\ and\ \bibinfo {author} {\bibfnamefont {M.}~\bibnamefont {Davy}},\ }\bibfield  {title} {\bibinfo {title} {Coherent {{Wave Control}} in {{Complex Media}} with {{Arbitrary Wavefronts}}},\ }\href {https://doi.org/10.1103/PhysRevLett.126.193903} {\bibfield  {journal} {\bibinfo  {journal} {Phys. Rev. Lett.}\ }\textbf {\bibinfo {volume} {126}},\ \bibinfo {pages} {193903} (\bibinfo {year} {2021}{\natexlab{a}})}\BibitemShut {NoStop}%
\bibitem [{\citenamefont {Bouchet}\ \emph {et~al.}(2021)\citenamefont {Bouchet}, \citenamefont {Rotter},\ and\ \citenamefont {Mosk}}]{bouchetMaximumInformationStates2021}%
  \BibitemOpen
  \bibfield  {author} {\bibinfo {author} {\bibfnamefont {D.}~\bibnamefont {Bouchet}}, \bibinfo {author} {\bibfnamefont {S.}~\bibnamefont {Rotter}},\ and\ \bibinfo {author} {\bibfnamefont {A.~P.}\ \bibnamefont {Mosk}},\ }\bibfield  {title} {\bibinfo {title} {Maximum information states for coherent scattering measurements},\ }\href {https://doi.org/10.1038/s41567-020-01137-4} {\bibfield  {journal} {\bibinfo  {journal} {Nat. Phys.}\ }\textbf {\bibinfo {volume} {17}},\ \bibinfo {pages} {564} (\bibinfo {year} {2021})}\BibitemShut {NoStop}%
\bibitem [{\citenamefont {Bouchet}\ and\ \citenamefont {Bossy}(2023)}]{bouchetTemporalShapingWave2023}%
  \BibitemOpen
  \bibfield  {author} {\bibinfo {author} {\bibfnamefont {D.}~\bibnamefont {Bouchet}}\ and\ \bibinfo {author} {\bibfnamefont {E.}~\bibnamefont {Bossy}},\ }\bibfield  {title} {\bibinfo {title} {Temporal shaping of wave fields for optimally precise measurements in scattering environments},\ }\href {https://doi.org/10.1103/PhysRevResearch.5.013144} {\bibfield  {journal} {\bibinfo  {journal} {Phys. Rev. Research}\ }\textbf {\bibinfo {volume} {5}},\ \bibinfo {pages} {013144} (\bibinfo {year} {2023})}\BibitemShut {NoStop}%
\bibitem [{\citenamefont {Yeo}\ \emph {et~al.}(2022)\citenamefont {Yeo}, \citenamefont {Leconte}, \citenamefont {{del Hougne}}, \citenamefont {Besnier},\ and\ \citenamefont {Davy}}]{yeoTimeReversalCommunications2022}%
  \BibitemOpen
  \bibfield  {author} {\bibinfo {author} {\bibfnamefont {K.~B.}\ \bibnamefont {Yeo}}, \bibinfo {author} {\bibfnamefont {C.}~\bibnamefont {Leconte}}, \bibinfo {author} {\bibfnamefont {P.}~\bibnamefont {{del Hougne}}}, \bibinfo {author} {\bibfnamefont {P.}~\bibnamefont {Besnier}},\ and\ \bibinfo {author} {\bibfnamefont {M.}~\bibnamefont {Davy}},\ }\bibfield  {title} {\bibinfo {title} {Time {{Reversal Communications With Channel State Information Estimated From Impedance Modulation}} at the {{Receiver}}},\ }\href {https://doi.org/10.1109/ACCESS.2022.3201559} {\bibfield  {journal} {\bibinfo  {journal} {IEEE Access}\ }\textbf {\bibinfo {volume} {10}},\ \bibinfo {pages} {91119} (\bibinfo {year} {2022})}\BibitemShut {NoStop}%
\bibitem [{\citenamefont {Sol}\ \emph {et~al.}(2024)\citenamefont {Sol}, \citenamefont {Magoarou},\ and\ \citenamefont {{del Hougne}}}]{solOptimalBlindFocusing2024}%
  \BibitemOpen
  \bibfield  {author} {\bibinfo {author} {\bibfnamefont {J.}~\bibnamefont {Sol}}, \bibinfo {author} {\bibfnamefont {L.~L.}\ \bibnamefont {Magoarou}},\ and\ \bibinfo {author} {\bibfnamefont {P.}~\bibnamefont {{del Hougne}}},\ }\href@noop {} {\bibinfo {title} {Optimal blind focusing on perturbation-inducing targets in sub-unitary complex media}} (\bibinfo {year} {2024}),\ \Eprint {https://arxiv.org/abs/2401.15415} {arxiv:2401.15415 [physics]} \BibitemShut {NoStop}%
\bibitem [{\citenamefont {Berkovits}(1991)}]{berkovitsSensitivityMultiplescatteringSpeckle1991}%
  \BibitemOpen
  \bibfield  {author} {\bibinfo {author} {\bibfnamefont {R.}~\bibnamefont {Berkovits}},\ }\bibfield  {title} {\bibinfo {title} {Sensitivity of the multiple-scattering speckle pattern to the motion of a single scatterer},\ }\href {https://doi.org/10.1103/PhysRevB.43.8638} {\bibfield  {journal} {\bibinfo  {journal} {Phys. Rev. B}\ }\textbf {\bibinfo {volume} {43}},\ \bibinfo {pages} {8638} (\bibinfo {year} {1991})}\BibitemShut {NoStop}%
\bibitem [{\citenamefont {Orazbayev}\ \emph {et~al.}(2024)\citenamefont {Orazbayev}, \citenamefont {Mall{\'e}jac}, \citenamefont {Bachelard}, \citenamefont {Rotter},\ and\ \citenamefont {Fleury}}]{orazbayev2024wave}%
  \BibitemOpen
  \bibfield  {author} {\bibinfo {author} {\bibfnamefont {B.}~\bibnamefont {Orazbayev}}, \bibinfo {author} {\bibfnamefont {M.}~\bibnamefont {Mall{\'e}jac}}, \bibinfo {author} {\bibfnamefont {N.}~\bibnamefont {Bachelard}}, \bibinfo {author} {\bibfnamefont {S.}~\bibnamefont {Rotter}},\ and\ \bibinfo {author} {\bibfnamefont {R.}~\bibnamefont {Fleury}},\ }\bibfield  {title} {\bibinfo {title} {Wave-momentum shaping for moving objects in heterogeneous and dynamic media},\ }\href@noop {} {\bibfield  {journal} {\bibinfo  {journal} {Nature Physics}\ ,\ \bibinfo {pages} {1}} (\bibinfo {year} {2024})}\BibitemShut {NoStop}%
\bibitem [{\citenamefont {Horodynski}\ \emph {et~al.}(2021)\citenamefont {Horodynski}, \citenamefont {Bouchet}, \citenamefont {K{\"u}hmayer},\ and\ \citenamefont {Rotter}}]{horodynskiInvariancePropertyFisher2021}%
  \BibitemOpen
  \bibfield  {author} {\bibinfo {author} {\bibfnamefont {M.}~\bibnamefont {Horodynski}}, \bibinfo {author} {\bibfnamefont {D.}~\bibnamefont {Bouchet}}, \bibinfo {author} {\bibfnamefont {M.}~\bibnamefont {K{\"u}hmayer}},\ and\ \bibinfo {author} {\bibfnamefont {S.}~\bibnamefont {Rotter}},\ }\bibfield  {title} {\bibinfo {title} {Invariance {{Property}} of the {{Fisher Information}} in {{Scattering Media}}},\ }\href {https://doi.org/10.1103/PhysRevLett.127.233201} {\bibfield  {journal} {\bibinfo  {journal} {Phys. Rev. Lett.}\ }\textbf {\bibinfo {volume} {127}},\ \bibinfo {pages} {233201} (\bibinfo {year} {2021})}\BibitemShut {NoStop}%
\bibitem [{\citenamefont {H{\"u}pfl}\ \emph {et~al.}(2024)\citenamefont {H{\"u}pfl}, \citenamefont {Russo}, \citenamefont {Rachbauer}, \citenamefont {Bouchet}, \citenamefont {Lu}, \citenamefont {Kuhl},\ and\ \citenamefont {Rotter}}]{hupflContinuityEquationFlow2024}%
  \BibitemOpen
  \bibfield  {author} {\bibinfo {author} {\bibfnamefont {J.}~\bibnamefont {H{\"u}pfl}}, \bibinfo {author} {\bibfnamefont {F.}~\bibnamefont {Russo}}, \bibinfo {author} {\bibfnamefont {L.~M.}\ \bibnamefont {Rachbauer}}, \bibinfo {author} {\bibfnamefont {D.}~\bibnamefont {Bouchet}}, \bibinfo {author} {\bibfnamefont {J.}~\bibnamefont {Lu}}, \bibinfo {author} {\bibfnamefont {U.}~\bibnamefont {Kuhl}},\ and\ \bibinfo {author} {\bibfnamefont {S.}~\bibnamefont {Rotter}},\ }\bibfield  {title} {\bibinfo {title} {Continuity equation for the flow of {{Fisher}} information in wave scattering},\ }\href {https://doi.org/10.1038/s41567-024-02519-8} {\bibfield  {journal} {\bibinfo  {journal} {Nat. Phys.}\ }\textbf {\bibinfo {volume} {20}},\ \bibinfo {pages} {1294} (\bibinfo {year} {2024})}\BibitemShut {NoStop}%
\bibitem [{\citenamefont {Fleming}\ \emph {et~al.}(2019)\citenamefont {Fleming}, \citenamefont {Conti},\ and\ \citenamefont {Di~Falco}}]{flemingPerturbationTransmissionMatrices2019}%
  \BibitemOpen
  \bibfield  {author} {\bibinfo {author} {\bibfnamefont {A.}~\bibnamefont {Fleming}}, \bibinfo {author} {\bibfnamefont {C.}~\bibnamefont {Conti}},\ and\ \bibinfo {author} {\bibfnamefont {A.}~\bibnamefont {Di~Falco}},\ }\bibfield  {title} {\bibinfo {title} {Perturbation of {{Transmission Matrices}} in {{Nonlinear Random Media}}},\ }\href {https://doi.org/10.1002/andp.201900091} {\bibfield  {journal} {\bibinfo  {journal} {Annalen der Physik}\ }\textbf {\bibinfo {volume} {531}},\ \bibinfo {pages} {1900091} (\bibinfo {year} {2019})}\BibitemShut {NoStop}%
\bibitem [{sup()}]{supp_mat}%
  \BibitemOpen
  \href@noop {} {\bibinfo {title} {See {{Supplemental Material}} at [{{URL}} will be inserted by publisher], which includes {Refs}.~[50--52], for the detailed theoretical analysis, additional simulations, and the phase correction method used for spatio-temporal focusing.}}\BibitemShut {Stop}%
\bibitem [{\citenamefont {Davy}\ \emph {et~al.}(2021)\citenamefont {Davy}, \citenamefont {K{\"u}hmayer}, \citenamefont {Gigan},\ and\ \citenamefont {Rotter}}]{davyMeanPathLength2021}%
  \BibitemOpen
  \bibfield  {author} {\bibinfo {author} {\bibfnamefont {M.}~\bibnamefont {Davy}}, \bibinfo {author} {\bibfnamefont {M.}~\bibnamefont {K{\"u}hmayer}}, \bibinfo {author} {\bibfnamefont {S.}~\bibnamefont {Gigan}},\ and\ \bibinfo {author} {\bibfnamefont {S.}~\bibnamefont {Rotter}},\ }\bibfield  {title} {\bibinfo {title} {Mean path length invariance in wave-scattering beyond the diffusive regime},\ }\href {https://doi.org/10.1038/s42005-021-00585-5} {\bibfield  {journal} {\bibinfo  {journal} {Commun Phys}\ }\textbf {\bibinfo {volume} {4}},\ \bibinfo {pages} {85} (\bibinfo {year} {2021})}\BibitemShut {NoStop}%
\bibitem [{\citenamefont {Del~Hougne}\ \emph {et~al.}(2021{\natexlab{b}})\citenamefont {Del~Hougne}, \citenamefont {Sobry}, \citenamefont {Legrand}, \citenamefont {Mortessagne}, \citenamefont {Kuhl},\ and\ \citenamefont {Davy}}]{delhougneExperimentalRealizationOptimal2021}%
  \BibitemOpen
  \bibfield  {author} {\bibinfo {author} {\bibfnamefont {P.}~\bibnamefont {Del~Hougne}}, \bibinfo {author} {\bibfnamefont {R.}~\bibnamefont {Sobry}}, \bibinfo {author} {\bibfnamefont {O.}~\bibnamefont {Legrand}}, \bibinfo {author} {\bibfnamefont {F.}~\bibnamefont {Mortessagne}}, \bibinfo {author} {\bibfnamefont {U.}~\bibnamefont {Kuhl}},\ and\ \bibinfo {author} {\bibfnamefont {M.}~\bibnamefont {Davy}},\ }\bibfield  {title} {\bibinfo {title} {Experimental {{Realization}} of {{Optimal Energy Storage}} in {{Resonators Embedded}} in {{Scattering Media}}},\ }\href {https://doi.org/10.1002/lpor.202000335} {\bibfield  {journal} {\bibinfo  {journal} {Laser \& Photonics Reviews}\ }\textbf {\bibinfo {volume} {15}},\ \bibinfo {pages} {2000335} (\bibinfo {year} {2021}{\natexlab{b}})}\BibitemShut {NoStop}%
\bibitem [{\citenamefont {Sounas}\ and\ \citenamefont {Al{\`u}}(2018)}]{sounasFundamentalBoundsOperation2018}%
  \BibitemOpen
  \bibfield  {author} {\bibinfo {author} {\bibfnamefont {D.~L.}\ \bibnamefont {Sounas}}\ and\ \bibinfo {author} {\bibfnamefont {A.}~\bibnamefont {Al{\`u}}},\ }\bibfield  {title} {\bibinfo {title} {Fundamental bounds on the operation of {{Fano}} nonlinear isolators},\ }\href {https://doi.org/10.1103/PhysRevB.97.115431} {\bibfield  {journal} {\bibinfo  {journal} {Phys. Rev. B}\ }\textbf {\bibinfo {volume} {97}},\ \bibinfo {pages} {115431} (\bibinfo {year} {2018})}\BibitemShut {NoStop}%
\bibitem [{\citenamefont {Cotrufo}\ \emph {et~al.}(2021)\citenamefont {Cotrufo}, \citenamefont {Mann}, \citenamefont {Moussa},\ and\ \citenamefont {Alu}}]{cotrufoNonlinearityinducedNonreciprocityPart2021}%
  \BibitemOpen
  \bibfield  {author} {\bibinfo {author} {\bibfnamefont {M.}~\bibnamefont {Cotrufo}}, \bibinfo {author} {\bibfnamefont {S.~A.}\ \bibnamefont {Mann}}, \bibinfo {author} {\bibfnamefont {H.}~\bibnamefont {Moussa}},\ and\ \bibinfo {author} {\bibfnamefont {A.}~\bibnamefont {Alu}},\ }\bibfield  {title} {\bibinfo {title} {Nonlinearity-{{Induced Nonreciprocity}}---{{Part I}}},\ }\href {https://doi.org/10.1109/TMTT.2021.3079250} {\bibfield  {journal} {\bibinfo  {journal} {IEEE Trans. Microwave Theory Techn.}\ }\textbf {\bibinfo {volume} {69}},\ \bibinfo {pages} {3569} (\bibinfo {year} {2021})}\BibitemShut {NoStop}%
\bibitem [{\citenamefont {Wang}\ \emph {et~al.}(2023)\citenamefont {Wang}, \citenamefont {Kononchuk}, \citenamefont {Kuhl},\ and\ \citenamefont {Kottos}}]{wangLossInducedViolationFundamental2023}%
  \BibitemOpen
  \bibfield  {author} {\bibinfo {author} {\bibfnamefont {C.-Z.}\ \bibnamefont {Wang}}, \bibinfo {author} {\bibfnamefont {R.}~\bibnamefont {Kononchuk}}, \bibinfo {author} {\bibfnamefont {U.}~\bibnamefont {Kuhl}},\ and\ \bibinfo {author} {\bibfnamefont {T.}~\bibnamefont {Kottos}},\ }\bibfield  {title} {\bibinfo {title} {Loss-{{Induced Violation}} of the {{Fundamental Transmittance-Asymmetry Bound}} in {{Nonlinear Complex Wave Systems}}},\ }\href {https://doi.org/10.1103/PhysRevLett.131.123801} {\bibfield  {journal} {\bibinfo  {journal} {Phys. Rev. Lett.}\ }\textbf {\bibinfo {volume} {131}},\ \bibinfo {pages} {123801} (\bibinfo {year} {2023})}\BibitemShut {NoStop}%
\bibitem [{\citenamefont {Del~Hougne}\ \emph {et~al.}(2016)\citenamefont {Del~Hougne}, \citenamefont {Lemoult}, \citenamefont {Fink},\ and\ \citenamefont {Lerosey}}]{delhougneSpatiotemporalWaveFront2016}%
  \BibitemOpen
  \bibfield  {author} {\bibinfo {author} {\bibfnamefont {P.}~\bibnamefont {Del~Hougne}}, \bibinfo {author} {\bibfnamefont {F.}~\bibnamefont {Lemoult}}, \bibinfo {author} {\bibfnamefont {M.}~\bibnamefont {Fink}},\ and\ \bibinfo {author} {\bibfnamefont {G.}~\bibnamefont {Lerosey}},\ }\bibfield  {title} {\bibinfo {title} {Spatiotemporal {{Wave Front Shaping}} in a {{Microwave Cavity}}},\ }\href {https://doi.org/10.1103/PhysRevLett.117.134302} {\bibfield  {journal} {\bibinfo  {journal} {Phys. Rev. Lett.}\ }\textbf {\bibinfo {volume} {117}},\ \bibinfo {pages} {134302} (\bibinfo {year} {2016})}\BibitemShut {NoStop}%
\bibitem [{\citenamefont {Chen}\ \emph {et~al.}(2020)\citenamefont {Chen}, \citenamefont {Kottos},\ and\ \citenamefont {Anlage}}]{chenPerfectAbsorptionComplex2020}%
  \BibitemOpen
  \bibfield  {author} {\bibinfo {author} {\bibfnamefont {L.}~\bibnamefont {Chen}}, \bibinfo {author} {\bibfnamefont {T.}~\bibnamefont {Kottos}},\ and\ \bibinfo {author} {\bibfnamefont {S.~M.}\ \bibnamefont {Anlage}},\ }\bibfield  {title} {\bibinfo {title} {Perfect absorption in complex scattering systems with or without hidden symmetries},\ }\href {https://doi.org/10.1038/s41467-020-19645-5} {\bibfield  {journal} {\bibinfo  {journal} {Nat Commun}\ }\textbf {\bibinfo {volume} {11}},\ \bibinfo {pages} {5826} (\bibinfo {year} {2020})}\BibitemShut {NoStop}%
\bibitem [{\citenamefont {Ho}\ \emph {et~al.}(2014)\citenamefont {Ho}, \citenamefont {Yeh}, \citenamefont {Neofytou}, \citenamefont {Kim}, \citenamefont {Tanabe}, \citenamefont {Patlolla}, \citenamefont {Beygui},\ and\ \citenamefont {Poon}}]{hoWirelessPowerTransfer2014}%
  \BibitemOpen
  \bibfield  {author} {\bibinfo {author} {\bibfnamefont {J.~S.}\ \bibnamefont {Ho}}, \bibinfo {author} {\bibfnamefont {A.~J.}\ \bibnamefont {Yeh}}, \bibinfo {author} {\bibfnamefont {E.}~\bibnamefont {Neofytou}}, \bibinfo {author} {\bibfnamefont {S.}~\bibnamefont {Kim}}, \bibinfo {author} {\bibfnamefont {Y.}~\bibnamefont {Tanabe}}, \bibinfo {author} {\bibfnamefont {B.}~\bibnamefont {Patlolla}}, \bibinfo {author} {\bibfnamefont {R.~E.}\ \bibnamefont {Beygui}},\ and\ \bibinfo {author} {\bibfnamefont {A.~S.~Y.}\ \bibnamefont {Poon}},\ }\bibfield  {title} {\bibinfo {title} {Wireless power transfer to deep-tissue microimplants},\ }\href {https://doi.org/10.1073/pnas.1403002111} {\bibfield  {journal} {\bibinfo  {journal} {Proc. Natl. Acad. Sci. U.S.A.}\ }\textbf {\bibinfo {volume} {111}},\ \bibinfo {pages} {7974} (\bibinfo {year} {2014})}\BibitemShut {NoStop}%
\bibitem [{\citenamefont {Yaghjian}(1980)}]{yaghjian_electric_1980}%
  \BibitemOpen
  \bibfield  {author} {\bibinfo {author} {\bibfnamefont {A.}~\bibnamefont {Yaghjian}},\ }\bibfield  {title} {\bibinfo {title} {Electric dyadic {Green}'s functions in the source region},\ }\href {https://doi.org/10.1109/PROC.1980.11620} {\bibfield  {journal} {\bibinfo  {journal} {Proceedings of the IEEE}\ }\textbf {\bibinfo {volume} {68}},\ \bibinfo {pages} {248} (\bibinfo {year} {1980})}\BibitemShut {NoStop}%
\bibitem [{\citenamefont {H\"upfl}\ \emph {et~al.}(2023)\citenamefont {H\"upfl}, \citenamefont {Bachelard}, \citenamefont {Kaczvinszki}, \citenamefont {Horodynski}, \citenamefont {K\"uhmayer},\ and\ \citenamefont {Rotter}}]{PhysRevA.107.023112}%
  \BibitemOpen
  \bibfield  {author} {\bibinfo {author} {\bibfnamefont {J.}~\bibnamefont {H\"upfl}}, \bibinfo {author} {\bibfnamefont {N.}~\bibnamefont {Bachelard}}, \bibinfo {author} {\bibfnamefont {M.}~\bibnamefont {Kaczvinszki}}, \bibinfo {author} {\bibfnamefont {M.}~\bibnamefont {Horodynski}}, \bibinfo {author} {\bibfnamefont {M.}~\bibnamefont {K\"uhmayer}},\ and\ \bibinfo {author} {\bibfnamefont {S.}~\bibnamefont {Rotter}},\ }\bibfield  {title} {\bibinfo {title} {Optimal cooling of multiple levitated particles: Theory of far-field wavefront shaping},\ }\href {https://doi.org/10.1103/PhysRevA.107.023112} {\bibfield  {journal} {\bibinfo  {journal} {Phys. Rev. A}\ }\textbf {\bibinfo {volume} {107}},\ \bibinfo {pages} {023112} (\bibinfo {year} {2023})}\BibitemShut {NoStop}%
\bibitem [{\citenamefont {Chong}\ and\ \citenamefont {Stone}(2011)}]{chongHiddenBlackCoherent2011}%
  \BibitemOpen
  \bibfield  {author} {\bibinfo {author} {\bibfnamefont {Y.~D.}\ \bibnamefont {Chong}}\ and\ \bibinfo {author} {\bibfnamefont {A.~D.}\ \bibnamefont {Stone}},\ }\bibfield  {title} {\bibinfo {title} {Hidden {{Black}}: {{Coherent Enhancement}} of {{Absorption}} in {{Strongly Scattering Media}}},\ }\href {https://doi.org/10.1103/PhysRevLett.107.163901} {\bibfield  {journal} {\bibinfo  {journal} {Phys. Rev. Lett.}\ }\textbf {\bibinfo {volume} {107}},\ \bibinfo {pages} {163901} (\bibinfo {year} {2011})}\BibitemShut {NoStop}%
\end{thebibliography}%

%%%%%%%%%% Merge with supplemental materials %%%%%%%%%%
\widetext
\begin{center}
\textbf{\large Supplemental Material for ``Detecting and Focusing on a Nonlinear Target in a Complex Medium"}
\end{center}
%%%%%%%%%% Merge with supplemental materials %%%%%%%%%%
%%%%%%%%%% Prefix a "S" to all equations, figures, tables and reset the counter %%%%%%%%%%
\setcounter{equation}{0}
\setcounter{figure}{0}
\setcounter{table}{0}
\setcounter{page}{1}
\makeatletter
\renewcommand{\theequation}{S\arabic{equation}}
\renewcommand{\thefigure}{S\arabic{figure}}
\renewcommand{\bibnumfmt}[1]{[S#1]}
\renewcommand{\citenumfont}[1]{S#1}
\setcounter{secnumdepth}{4}
%%%%%%%%%% Prefix a "S" to all equations, figures, tables and reset the counter %%%%%%%%%%

\section{Theory}

We consider a system without free charges and no magnetisation so that the wave equation is given by
\begin{equation}
    \nabla \times (\nabla \times \vec{E}) = - \mu_0 \partial_t^2 \vec{D}.
\end{equation}
We can now split up the electric displacement field $\vec{D} = \epsilon_0 \epsilon \vec{E} + \vec{P}^{\text{NL}}$ into a linear part given by the relative electric permittivity $\epsilon$ and a polarization part $\vec{P}^{\text{NL}}[\vec{E}]$, which is nonlinear in $\vec{E}$. We are now going to make use of the Green's tensor $G_\omega$ of the linear system at frequency $\omega$ with outgoing boundary conditions, i.e.
\begin{equation}
    \nabla \times (\nabla \times G_\omega)(\vec{r},\vec{r}\,') - \frac{\omega^2}{c^2}  \epsilon(\vec{r}) G_\omega(\vec{r},\vec{r}\,') = \matrix{1} \delta(\vec{r}-\vec{r}\,').
\end{equation}
{This definition allows us to rewrite the monochromatic solution of the wave equation at frequency $\omega$ using the Green's tensor in the far field
\begin{equation}
\label{eq:sol_green}
    \vec{E}_\omega(\vec{r}) = \vec{E}_{\omega}^\text{L}(\vec{r}) + \frac{k^2}{\epsilon_0} \int G_\omega(\vec{r},\vec{r}\,') \vec{P}^{\text{NL}}_\omega[\vec{E}(\vec{r}\,')] d \vec{r}\,',
\end{equation}
where $k=\omega/c$, $\vec{E}_{\omega}^\text{L}$ describes the incident field component at frequency $\omega$ scattered in the linear system (i.e. $\nabla \times (\nabla \times \vec{E}_{\omega}^\text{L}) + k^2 \epsilon \vec{E}_{\omega}^\text{L} = 0$) and $\vec{P}_\omega^{\text{NL}}(\vec{E})$ the nonlinear polarization field at frequency $\omega$ corresponding to the full electric field $\vec{E}$. We want to emphasize here that $\vec{P}_\omega^{\text{NL}}(\vec{E})$ depends in general on the full electromagnetic field $\vec{E}$ at all frequencies and not just the field component $\vec{E}_{\omega}$ at frequency $\omega$. This occurs in cases such as second harmonic generation, where the polarization field $P_{2\omega}$ at $2 \omega$ depends on the electric field component $\vec{E}_\omega$ at $\omega$. }

\subsection{Separating the linear and nonlinear contributions}
Our first goal is going to be to separate these two contributions. For this we can make us of the fact that the linear scaling of the output field with respect to the input field is broken due to the presence of the nonlinearity. We proceed by varying the strength (quantified by $\alpha$) of the incident field $\vec{E}_{\omega}^{\text{L},\text{in}} \rightarrow \vec{E}_{\omega}^{\text{L},\text{in}}\alpha$, which produces a predictable linear response in the linear contributions of the scattered field $\vec{E}_\omega^{\text{L},\alpha} = \vec{E}_{\omega}^\text{L} \alpha$.
{Therefore by varying $\alpha$ we can extract the nonlinear contributions of the field by removing exactly those contributions with linear scaling
\begin{equation}
    \vec{E}_{\omega}^{\alpha+\Delta \alpha}- \frac{\alpha + \Delta \alpha}{\alpha} \vec{E}_{\omega}^{\alpha}  =  \frac{k^2}{\epsilon_0} \int G_\omega(\vec{r},\vec{r}') \left\{\vec{P}_{\omega}^{\text{NL}}[\vec{E}^{\alpha+\Delta \alpha}(\vec{r}\,')] - \frac{\alpha + \Delta \alpha}{\alpha} \vec{P}_{\omega}^{\text{NL}}[\vec{E}^{\alpha}(\vec{r}\,')] \right\} d \vec{r}\,'
\end{equation}
whereby the superscript $\alpha,\alpha+\Delta \alpha$ quantifies the strength of the incident field.
The left hand side of the equation now effectively corresponds to a field that originates at the nonlinearity and propagates out of the system according to the Green's tensor of the corresponding linear system $G_\omega(\vec{r},\vec{r}')$. This already shows that a nonlinearity can be detected just by observing the divergence of the outgoing field from the linear field strength scaling property. In order to rewrite this expression to Eq.~(2) of the main text we divide by $\alpha + \Delta \alpha$ and define the difference of the normalized fields $\delta \vec{E}_{\omega}(\vec{r}) = (\alpha+\Delta \alpha)^{-1} \vec{E}_{\omega}^{\alpha+\Delta \alpha}(\vec{r}) - \alpha^{-1} \vec{E}_{\omega}^\alpha(\vec{r})$, which satisfies 
\begin{equation}
\label{eq:Coupling_Greens}
    \delta \vec{E}_{\omega}(\vec{r}) = \frac{k^2}{\epsilon_0} \int d\vec{r}\,' G_{\omega}(\vec{r},\vec{r}\,') \delta \vec{P}_{\omega}^{\mathrm{NL}}(\vec{r}\,')\,,
\end{equation}
\noindent with $\delta \vec{P}_{\omega}^{\mathrm{NL}}(\vec{r}\,') = (\alpha + \Delta \alpha)^{-1} \vec{P}_{\omega}^{\mathrm{NL}}[\vec{E}^{\alpha + \Delta \alpha}(\vec{r}\,')]- \alpha^{-1} \vec{P}_{\omega}^{\mathrm{NL}}[\vec{E}^\alpha(\vec{r}\,')]$. In the next section we are going to see how this can be exploited to create a focusing field on a point-like target.}

\subsection{Focusing on the nonlinearity}
\label{sec:optimal_focusing}
Our next objective will be to find an incident wavefront that focuses onto the location of the nonlinearity. We reverse the outgoing finite difference field (assumed non-zero from here on) derived in the previous section so that the incident field is given by $\vec{E}_\omega^\text{opt} = \delta \vec{E}_\omega^*$ in the far field and creates a focus at the nonlinear dielectric.
{This can be seen for systems, where $\epsilon$ is real valued (i.e. no absorption or gain present) by using Green's identity to identify the linear response $\vec{E}_{\omega}^{\text{L},\text{opt}}$ of the system to the incident field $\delta \vec{E}_\omega^*$, i.e. a system with the nonlinearity removed,
\begin{equation}
    \vec{E}_{\omega}^{\text{L},\text{opt}}(\vec{r}_1) =  \int_{S^3_R}  [\vec{n} \times G_\omega(\vec{r},\vec{r}_1)] \cdot [\nabla \times \vec{E}_{\omega}^{\text{L},\text{opt}}] - [\nabla \times G_\omega(\vec{r},\vec{r}_1)] \cdot [\vec{n} \times \vec{E}_{\omega}^{\text{L},\text{opt}}]   d \sigma,
\end{equation}
where $S^3_R$ is the surface of a sphere with large radius $R$ centered at the origin. Due to the Green's function having outgoing boundary conditions (Silver-Müller radiation condition), the outgoing part of $\vec{E}_{\omega}^{\text{L},\text{opt}}$ does not contribute to this equation and we thus get
\begin{equation}
\label{eq:Green_E}
    \vec{E}_{\omega}^{\text{L},\text{opt}}(\vec{r}_1) =  \int_{S^3_R}  [\vec{n} \times G_\omega(\vec{r},\vec{r}_1)] \cdot [\nabla \times \delta \vec{E}^*_\omega(\vec{r})] - [\nabla \times G_\omega(\vec{r},\vec{r}_1)] \cdot [\vec{n} \times \delta \vec{E}^*_\omega(\vec{r})]   d \sigma.
\end{equation}}
We can simplify this expression by using the Green's function identity
\begin{equation}
\label{eq:Green_identity}
\begin{split}
    - 2 i \Im (G_\omega (\vec{r}_0,\vec{r}_1)) =& -2 i \int_{S^3_R}  (\nabla \times \Im (G_\omega)(\vec{r},\vec{r}_1)) \cdot (\vec{n} \times G_\omega(\vec{r},\vec{r}_0))  - (\vec{n} \times \Im (G_\omega)(\vec{r},\vec{r}_1))\cdot (\nabla \times G_\omega(\vec{r},\vec{r}_0))  d \sigma\\
    =& \int_{S^3_R}  (\nabla \times G_\omega^*(\vec{r},\vec{r}_1)) \cdot (\vec{n} \times G_\omega(\vec{r},\vec{r}_0))  - (\vec{n} \times G_\omega^*(\vec{r},\vec{r}_1))\cdot (\nabla \times G_\omega(\vec{r},\vec{r}_0))  d \sigma.
\end{split}
\end{equation}
The second equality holds due to the outgoing boundary condition of $G_\omega$. Note that we avoided the ambiguity of the dyadic Green's function in the source region \cite{yaghjian_electric_1980} in Eq.~\eqref{eq:Green_E}/\eqref{eq:Green_identity} because $\vec{E}_{\omega}^{\text{L},\text{opt}},\Im(G_\omega)$ are solutions of the source free wave equation inside the system.

By combining Eq.~\eqref{eq:Coupling_Greens} with \eqref{eq:Green_E} and \eqref{eq:Green_identity} this gives us the linear contribution of the field at the nonlinearities
\begin{equation}
\label{eq:focusing_lin}
    \vec{E}_{\omega}^{\text{L},\text{opt}}(\vec{r}) = - \frac{ 2 i k^2}{ \epsilon_0} \int \Im G_\omega(\vec{r},\vec{r}\,') (\delta \vec{P}_{\omega}^{\text{NL}}(\vec{r}\,'))^*  d \vec{r}\,'.
\end{equation}
\rev{Since we assumed that the probing field produces a nonlinear response, i.e., $\delta \vec{E}_\omega \neq 0$, it follows directly from the definition that $\delta \vec{P}_{\omega}^{\text{NL}}(\vec{r},')$ must also be non-zero. Eq.~\eqref{eq:focusing_lin} therefore implies that the electric field within the nonlinear region $\vec{E}_{\omega}^{\text{L},\text{opt}}(\vec{r})$ can be expected to be non-zero.}
%However, in this general formulation it is not possible to show based on eq.~\eqref{eq:focusing_lin} that the focusing on the non-linearities is optimal. 

%On the one hand this formula indicates that we can expect a field inside of the nonlinearity.
%However, on the other hand, the term $\Im G_\omega(\vec{r},\vec{r}\,')$ can deteriorate the focusing on the nonlinearity and we have so far not included the nonlinear response of the target, which will in general introduce difficulties in predicting the exact field. 

Overall for general nonlinearities (e.g. multiple nonlinearities, extended nonlinearities in space) our method is able to identify incident states that focus at the spatial region, where these non-linearities are located. 
\rev{However, both the spatial correlations of the electromagnetic field, captured by $\Im\  G_\omega(\vec{r},\vec{r},')$, and the nonlinear response of the target, $\delta \vec{P}_{\omega}^{\text{NL}}$, obfuscate the exact field intensity at the target. This highlights the need for additional system information, such as details of the nonlinearities in the system, in order to find the optimal input wavefront for focusing.} 
In our case we overcome this by considering systems with only a single input mode that couples to the nonlinearity such as in the case of a nonlinearity made up of an antenna measuring only a singular polarization direction connected to a LNA or a point-like nonlinearity in restricted systems with only a singular polarization degree of freedom.

\subsection{Point-like nonlinearity}
We consider a point-like nonlinearity in the Rayleigh regime with center at $\vec{r}_0$ volume $V_R$ and diameter $R$, so that the far field is given by
\begin{equation}
\label{eq:pointlike}
\vec{E}_\omega(\vec{r}) = \vec{E}_{\omega}^\text{L}(\vec{r}) + \frac{k^2 V_R}{\epsilon_0}  G_\omega(\vec{r},\vec{r}_0) \vec{P}^{\text{NL}}_\omega[\vec{E}(\vec{r}_0)].
\end{equation}
Thus the outgoing field pattern is only given by the Green's tensor of the linear system $G_\omega(\vec{r},\vec{r}_0)$ coupling the location of the nonlinearity to the far field. Based on this we can simplify Eq.~\eqref{eq:focusing_lin} giving us the linear contribution
\begin{equation}
    \vec{E}_{\omega}^{\text{L},\text{opt}}(\vec{r}_0) = - \Im G_\omega(\vec{r}_0,\vec{r}_0) \frac{ 2 i k^2 V_R}{\epsilon_0} (\delta \vec{P}_{\omega}^{\text{NL}}(\vec{r}_0))^* .
\end{equation}
Thus in the linear reference system a focus can be created at the location of the nonlinearity. {In the case, where $\vec{P}_{\omega}^{\text{NL}}$ is restricted to a singular polarization direction (e.g. antenna) then only a singular input wavefront can focus onto our target and the focus is therefore optimal.}

\section{Far field description}
\label{sec:farfield}
{We consider the case of waves at the frequency $\omega$ and a basis of the wavefronts in the far field $\vec{E}^n_\omega$ so that the in- and outgoing fields have the corresponding coefficients $\vec{c}^\text{in},\vec{c}^\text{out}$ (see section \ref{sec:Smatrix}).We will for now focus on the case, where the nonlinearity only couples to the system through a single polarization degree of freedom (e.g. an antenna or a system allowing only a single polarization direction), while the more general case is considered in section \ref{sec:extended}. In linear systems these coefficients can be connected by the linear scattering matrix $\matrix{S}^\text{L}$. However in our case this does no longer hold true due to the nonlinearity violating the superposition principle, where the connection $\vec{c}^\text{out} = \hat{S}(\vec{c}^\text{in})$ is given by the non-linear scattering operator $\hat{S}$. 
Luckily, those incident fields that do not interact with the nonlinearity stay linear, which means that for a point-like nonlinearity an incident field quantified by $\vec{d}^*$ can be found, so that all orthogonal incident fields do not interact with the nonlinearity, i.e.
\begin{equation}
        \hat{S}(\vec{c}^\text{in}) - \matrix{S}^\text{L}\vec{c}^\text{in} = 0
\end{equation}
whereby $\matrix{S}^\text{L}$ is the scattering matrix of the linear system without the nonlinearity. This can be seen by considering that the vector $\vec{d}$ corresponds to the far field coefficients of $G_\omega(\vec{r},\vec{r}_0)$ in our chosen basis, i.e. $d_j = \frac{- i}{2 \omega \mu_0} \int_{S^3_R} (\vec{n} \times (\vec{E}_\omega^j)(r)) \cdot (\nabla \times G_{\omega}(r,r_0)) -  (\nabla \times (\vec{E}_\omega^j)(r)) \cdot (\vec{n} \times G_{\omega}(r,r_0)) d \sigma$ (see section \ref{sec:Smatrix}). Using Green's identity we can see that $d_j = \frac{i}{2 \omega \mu_0} (\vec{E}_\omega^{\text{L},j})(r_0)$, where $\vec{E}_\omega^{\text{L},j}$ corresponds to the solution of the linear system of the far field mode $j$. Thus if we take an incident field in the input channels $\vec{c}^\text{in} \cdot \vec{d} = 0$ then the linear part of the field is given by $\vec{E}_\omega^\text{L}(\vec{r}_0) = \sum_j c_j \vec{E}_\omega^{\text{L},j}(\vec{r}_0) \propto \vec{d} \cdot \vec{c} = 0$, showing that no field is present at the location of the nonlinearity. Due to the non-linear polarization $P^\text{NL}$ being only non-zero for electromagnetic fields that interact with the non-linearity ($E(r_0) \neq 0$), this means that only the components of $\vec{c}^\text{in}$ parallel to $\vec{d}^*$ result in differences between $\hat{S}$ and $\matrix{S}^\text{L}$, i.e.
\begin{equation}
    \hat{S}(\vec{c}^\text{in}) - \matrix{S}^\text{L}\vec{c}^\text{in} = \hat{S}(\hat{P}_{\vec{d}^*}\vec{c}^\text{in}) - \matrix{S}^\text{L}\hat{P}_{\vec{d}^*}\vec{c}^\text{in},
\end{equation}
where $\hat{P}_{\vec{d}^*} = \vec{d}^* \vec{d}^T / \vert \vec{d} \vert^2$ is the projection operator onto the vector $\vec{d}^*$.
We can simplify this even further by noting that Eq.~\eqref{eq:pointlike} tells us that the part of the outgoing field due to the non-linearity results in an outgoing field given by $G_\omega(\vec{r},\vec{r}_0)$, which is described by $\vec{d}$ so that we get
\begin{equation}
    \hat{S}(\vec{c}^\text{in}) - \matrix{S}^\text{L}\vec{c}^\text{in} = \hat{P}_{\vec{d}}\left(\hat{S}(\hat{P}_{\vec{d}^*}\vec{c}^\text{in}) - \matrix{S}^\text{L}\hat{P}_{\vec{d}^*}\vec{c}^\text{in}\right),
\end{equation}
where $\hat{P}_{\vec{d}}$ is the projection operator onto the vector $\vec{d}$.
By inserting the definition of the orthogonal projection operators we can now describe the scattering of the nonlinear scattering operator by
\begin{equation}
    \hat{S}(\vec{c}^\text{in}) = \matrix{S}^\text{L} \vec{c}^\text{in} + \frac{\vec{d} \vec{d}^\dagger}{\vert \vec{d} \vert^2}\left(\hat{S}\left(\frac{\vec{d}^* \vec{d}^T\vec{c}^\text{in}}{\vert \vec{d} \vert^2} \right) - \matrix{S}^\text{L}\frac{\vec{d}^* \vec{d}^T\vec{c}^\text{in}}{\vert \vec{d} \vert^2}\right)  = \matrix{S}^\text{L} \vec{c}^\text{in} + \vec{d} f(\vec{d} \cdot \vec{c}^\text{in}),
\end{equation}
whereby $\matrix{S}^\text{L}$ is the scattering matrix of the linear system without the nonlinearity and $f$ is the nonlinear scalar function describing the deviation of $\hat{S}$ from a linear relation, i.e. $f(x) = \vec{d}^\dagger[\hat{S}(\vec{d}^* x / \vert \vec{d}\vert^2) - S^L \vec{d}^* x / \vert \vec{d}\vert^2 ] / \vert \vec{d}\vert^2$.} This nonlinear operator can now be probed using a basis of incident channels, which can be conveniently summarized in the matrix $\matrix{S}^\alpha$ so that we have
\begin{equation}
\label{eq:Sop_matrix_rep}
    \matrix{S}^\alpha_{m,n}  = \matrix{S}^\text{L}_{m,n} + d_m \alpha^{-1} f(\alpha d_n),
\end{equation}
where we choose the basis of incident channels $\vec{c}^\text{in} = \alpha \vec{e}_n$ for all $n$, quantify the strength of the incident fields by $\alpha$ and normalize the outgoing fields by $\alpha$. This has the advantage that for linear system this reduces to the scattering matrix, while we can probe the nonlinear contributions of the scattering operator by varying $\alpha$. Note however that $\matrix{S}^\alpha$ does not fully describe the operator $\hat{S}$, due to the loss of the superposition principle in nonlinear systems, but only the response of the system for a basis of incident channels.

{\subsection{Connecting the far field patterns with the scattering matrix}
\label{sec:Smatrix}
We connect the scattering matrix with the far field pattern using the hermitian form based on the poynting vector \cite{PhysRevA.107.023112}
\begin{equation}
    \langle E_{\omega}^1, E_{\omega}^2 \rangle = \frac{1}{2} \int_{S^3_R} [(E_{\omega}^1)^* \times H_{\omega}^2 -  (H_{\omega}^1)^* \times E_{\omega}^2] \cdot  d \vec{n},
\end{equation}
where $S^3_R$ is the surface of a sphere with large radius $R$ centered at the origin. We can now use this product to define a basis of incident states $\vec{E}_\omega^j$ with power normalization, i.e.
\begin{equation}
    \langle E_{\omega}^i, E_{\omega}^j \rangle = - \delta_{i,j}.
\end{equation}
We now define the input coefficients $c^\text{in}$ corresponding to a field $E_\omega$ by
\begin{equation}
    c^\text{in}_i = - \langle E_{\omega}^i, E_{\omega} \rangle.
\end{equation}
Similarly the outgoing field if given by $(E_{\omega}^j)^*$, which can be used to define the outgoing coefficients $c^\text{out}$ by
\begin{equation}
    c^\text{out}_i = \langle (E_{\omega}^i)^*, E_{\omega} \rangle.
\end{equation}
The scattering operator is now given by the relation $c^\text{out} = \hat{S}[c^\text{in}]$, which is linear in the case where no non-linearity is present.
Finally, we can define the outgoing coefficients $\vec{d}$ corresponding to the Green's function (e.g. $E_\omega(r) = G_\omega(r,r_0)$, $H_\omega = \nabla \times E_\omega/(i \mu_0 \omega)$) by
\begin{equation}
    d_i = \langle (E_{\omega}^i)^*, G_\omega(\cdot,r_0) \rangle = \frac{- i}{2 \omega \mu_0} \int_{S^3_R} (\vec{n} \times (\vec{E}_\omega^j)(r)) \cdot (\nabla \times G_{\omega}(r,r_0)) -  (\nabla \times (\vec{E}_\omega^j)(r)) \cdot (\vec{n} \times G_{\omega}(r,r_0)) d \sigma.
\end{equation}}

\subsection{Extracting the focusing field}
We will now use Eq.~\eqref{eq:Sop_matrix_rep} at two incident powers $\alpha,\alpha+\Delta \alpha$ to define a matrix version of Eq.~\eqref{eq:Coupling_Greens}. The results are summarized by the matrices $\matrix{S}^\alpha,\matrix{S}^{\alpha+\Delta \alpha}$, which gives us
\begin{equation}
    \matrix{\Delta S}_{m,n} = (\matrix{S}^{\alpha+\Delta \alpha} - \matrix{S}^{\alpha} )_{m,n}  =  d_m \left((\alpha+\Delta \alpha)^{-1} f((\alpha+\Delta \alpha) d_n) - \alpha^{-1} f(\alpha d_n) \right).
\end{equation}
This rank-one matrix contains information of at least one incident channel that interacts with the nonlinearity, allowing us to apply a singular value decomposition on $\matrix{\Delta S}$ to extract $\vec{d}$. 
 Note that if multiple far field modes couple to the nonlinearity (e.g. non-scalar waves, multiple polarization degrees of freedom) or if multiple nonlinearities are present then the non-linearity acts on the subspace of these modes. In this case $f$ would need to be replaced by a function describing the nonlinear interaction between these modes and the rank of the matrix $\matrix{\Delta S}$ will be the dimension of this subspace (see section \ref{sec:extended}). In general while we can still use $\matrix{\Delta S}$ to identify modes that focus on the nonlinearity, in order to create an optimal focus more information on the non-linearity is needed.

\subsection{Reciprocity and nonlinearity detection}
One important property in most linear systems is the reciprocity condition, which can now break due to nonlinear interactions. In our case this turns out to be
\begin{equation}
    (\matrix{S}^\alpha - (\matrix{S}^\alpha)^T)_{m,n} = d_m  \alpha^{-1}f(\alpha d_n) - d_n \alpha^{-1} f(\alpha d_m).
\end{equation}
While in theory it is not guaranteed that we will see reciprocity breaking using $\matrix{S}^\alpha$ (e.g. if $\vec{d}$ corresponds to a basis vector $d_n \propto \delta_{m,n}$), in practice we saw that this can serve as a useful tool for the detection of the nonlinearity. 

\subsection{Lowest singular value and absorption}
We find a strong average correlation  between the wavefront for maximal focusing $\vec{c}^\text{in} = \vec{d}^*$ and the eigenvector $\vec{U}_N$ corresponding to the smallest eigenvalue $\sigma_N$ of $(S^\alpha)^\dagger S^\alpha$ that gives minimal reflection from the cavity and therefore maximal absorption~\cite{chongHiddenBlackCoherent2011}. The correlation coefficient averaged over the frequency range is $\langle |\vec{U}_N^* \cdot \vec{c}^\text{in}| \rangle \sim 0.8$. As the cavity is closed, absorption at the target is indeed the main loss mechanism. The correlation coefficient is however below unity since other dissipative mechanisms such as uniform absorption within the cavity also takes place.  Small eigenvalues $\sigma_N \rightarrow 0$ indicate that the incident energy is almost completely dissipated within the target.   

{
\subsection{Extended and multiple nonlinearities} \label{sec:extended}
We will now consider extended and multiple nonlinearities $D_i$. Similar to section \ref{sec:farfield} the far field modes that couple to the regions of these nonlinearities are spanned by $G_\omega(\vec{r},\vec{r}_0)$ for $\vec{r}_0 \in D = \cup_i D_i$, which we will describe in the far field by the basis $\vec{d}^1,\vec{d}^2,\dots,\vec{d}^N$ of dimension $N$. With the same arguments as in section \ref{sec:farfield} we can now show that for an incident field quantified by $\vec{c}$ and $\vec{c} \cdot \vec{d}^i = 0$ for all $i$, the corresponding electric field in the linear system disappears at the locations of the nonlinearities, i.e. $E^L_\omega(r_0) = 0$ for $\vec{r}_0 \in D$. As such the vectors $(\vec{d}^i)^*$ span the space of input modes that can couple to the nonlinearity and $\vec{d}^i$ the outgoing modes that the nonlinearity couples to. Using this we now write the scattering operator 
\begin{equation}
    \hat{S}[\vec{c}^\text{in}] = \matrix{S}^\text{L} \vec{c}^\text{in} + \sum_{i=1}^N \vec{d}^i f_i(\vec{d}^1 \cdot \vec{c}^\text{in},\dots,\vec{d}^N \cdot \vec{c}^\text{in}),
\end{equation}
for the non-linear functions $f_i$. Next we consider the difference matrix given by
\begin{equation}
    \matrix{\Delta S}_{m,n} = \sum_i d_m^i \left((\alpha+\Delta \alpha)^{-1} f_i((\alpha+\Delta \alpha) d_n^1,\dots,(\alpha+\Delta \alpha) d_n^N) - \alpha^{-1} f_i(\alpha d_n^1,\dots,\alpha d_n^N) \right).
\end{equation}
We can see that the dimension of $\matrix{\Delta S}$ is at most the number of in coupling modes $N$ and that the space of left singular values is spanned by $\vec{d}^i$. This shows that we can use $\matrix{\Delta S}$ to extract the incident wavefront that focus on the nonlinearities. However it is important to note that in general the vectors $\vec{d}^i$ will not be the set of left singular vectors, due to $\left((\alpha+\Delta \alpha)^{-1} f_i((\alpha+\Delta \alpha) d_n^1,\dots,(\alpha+\Delta \alpha) d_n^N) - \alpha^{-1} f_i(\alpha d_n^1,\dots,\alpha d_n^N) \right)$ being in general not orthogonal for different $i$. }

% \paragraph{Multiple noninteracting pointlike particles}
% If we have multiple pointlike nonlinearities, where we can neglect their interaction so that the individual nonlinearities do not interfere. Then we get 
% \begin{equation}
%     f_i(\vec{d}^1 \cdot \vec{c}^\text{in},\dots,\vec{d}^N \cdot \vec{c}^\text{in}) = f_i(\vec{d}^i \cdot \vec{c}^\text{in}).
% \end{equation}
% TODO

\begin{figure*}
\centering
\includegraphics[width=\columnwidth]{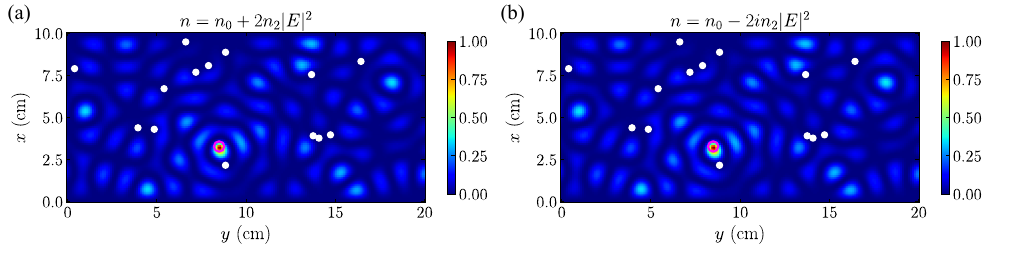}
\caption{\label{fig: comsol 2D 1 target} (a, b) Intensity map within the system for $\vec{c}^\text{in} = \vec{d}^*$ for a nonlinearity which intensity-dependent index is given by Eq.~\eqref{eq: n simu comsol}. Panel (a) corresponds to case 1 and panel (b) to case 2.  Both maps are normalized by the maximum value obtained for the optimized wavefront. The white dots represent the location of the metallic scatterers and the \rev{pink} circle the location of the nonlinear target.
}
\end{figure*}

{
\section{Simulations for controlled nonlinear behavior}
We present the results of numerical simulations obtained with COMSOL. The goal here is to demonstrate the versatility of the method with respect to the type of nonlinearity present in the system. We consider a 2D cavity with 9 antennas on each side at a frequency $f = 13.51$ GHz. The field inside the cavity is randomized through the presence of 14 randomly located metallic scatterers of 1 mm radius. The nonlinear target is defined as a dielectric scatterer with a 1 mm radius and Kerr-like index given by:
\begin{equation}\label{eq: n simu comsol}
    n = 
    \begin{cases}
        n_0 + n_2 |\vec{E}_{\omega}|^2 \quad \text{case 1}\\
        n_0 - i n_2 |\vec{E}_{\omega}|^2 \quad \text{case 2},
    \end{cases}
\end{equation}
where the case 1 corresponds to nonlinear dispersion, and the case 2 corresponds to nonlinear absorption. The intensity maps obtained from the first left singular vectors of $\Delta S$ in each cases (as described in the main text) are shown in Fig.~\ref{fig: comsol 2D 1 target}. The same locations are used for the scatterers and the target inside the system, and for the value of the nonlinear response $n_2 = 5\cdot 10^{-6}$ m$^2$/V$^2$. In both cases the incident wavefront obtained from the SVD focuses on the target \rev{regardless of the chosen nonlinear function, showing that the approach is independent of the type of nonlinearity}.}

\section{Correction of the phase for broadband signals}
The wavefront for optimal focusing in space and time corresponds to the time-reversed (or equivalently phase-conjugate) of the transmission coefficient $\vec{c}^\text{opt}(\omega) = \vec{t}^*(\omega)$. The phases at each frequency are aligned at the focus, meaning that the scattered wavefront $\vec{c}^\text{opt}(\omega) = \Delta S(\omega) \vec{c}^\text{opt}(\omega)$ acquires a phase equal to $\mathrm{arg}[t(\omega)]$. We therefore determine the phase $\phi(\omega) $ from the condition $\arg \left[ \vec{c}^\text{in}e^{i\phi} \Delta S \vec{c}^\text{in}e^{i\phi} \right] =0$. Because both $\vec{c}^\text{in}e^{i\phi}$ and $\vec{c}^\text{in}e^{i(\phi+\pi)}$ satisfy this condition, we finally exploit the continuity of $\phi(\omega)$ over the bandwidth to correct $\pi$-phase shifts.

%TC:endignore
\end{document}